\documentclass[prb,showpacs,aps,notitlepage,preprint]{revtex4}
\usepackage{amsmath}
\usepackage{epsfig}
\usepackage{mathdots}
\usepackage{graphicx}
\usepackage[ps2pdf,linkbordercolor={0 0 1}]{hyperref}
\newlength{\upit}\upit=0.1truein
\newcommand{\zmatrix}[4]{\left[\begin{matrix}#1 & #2\cr #3&#4\end{matrix}\right]}
\newcommand{\cmatrix}[4]{\left (\begin{matrix}#1 & #2\cr #3&#4\end{matrix}\right)}

\newcommand{\ltappr}{{{\lower4pthbox{$<$} } \atop \widetilde{ \ \ \ }}}
\newlength{\bxwidth}\bxwidth=1.5 truein

\begin{document}
\newcommand{\dg}{^{\dagger }}
\newcommand{\vk}{\vec k}
\newcommand{\vq}{{\vec{q}}}
\newcommand{\vp}{\bf{p}}
\newcommand{\bfk}{\mathbf k}
\newcommand{\al}{\alpha}
\newcommand{\be}{\beta}
\newcommand{\si}{\sigma}
\newcommand{\rarrow}{\rightarrow}
\def\fig#1#2{\includegraphics[height=#1]{#2}}
\def\figx#1#2{\includegraphics[width=#1]{#2}}
\newlength{\figwidth}
\figwidth=10cm
\newlength{\shift}
\shift=-0.2cm
\newcommand{\fg}[3]
{
\begin{figure}[ht]

\vspace*{-0cm}
\[
\includegraphics[width=\figwidth]{#1}
\]
\vskip -0.2cm
\caption{\label{#2}
\small#3
}
\end{figure}}
\newcommand{\fgb}[3]
{
\begin{figure}[b]
\vskip 0.0cm
\begin{equation}\label{}
\includegraphics[width=\figwidth]{#1}
\end{equation}
\vskip -0.2cm
\caption{\label{#2}
\small#3
}
\end{figure}}

\newcommand \bea {\begin{eqnarray} }
\newcommand \eea {\end{eqnarray}}
\newcommand \beg {\begin{equation} }
\newcommand \en {\end{equation}}
\newcommand{\bfp}{\mathbf p}
\newcommand{\bfq}{\mathbf q}
\newcommand{\bk}{{\bf{k}}}
\newcommand{\bx}{{\bf{x}}}
\newcommand{\bR}{{\bf{R}}}
\newcommand{\pu}{PuCoGa_{5}}
\newcommand{\np}{NpPd_{5}Al_{2}}
\newcommand{\bsig}{\mbox{\boldmath $\sigma$}}
\newcommand{\bS}{{\bf S}}
\newcommand{\bPhi}{\hat  \Phi}
\newcommand{\half}{{\textstyle \frac{1}{2}}}
\newcommand{\Nhalf}{{\textstyle \frac{N}{2}}}
\newcommand{\qter}{{\textstyle \frac{1}{4}}}

\title{Supplementary Material}


\centerline{\Large\bf   
Supplementary Information}


\vskip 0.3in
\centerline{\bf Online material: Heavy electrons and the symplectic symmetry of spin
}
\vskip 0.4in
\centerline{Rebecca Flint, M. Dzero and P. Coleman}
\centerline{\sl 
Center for Materials Theory,
Rutgers University, Piscataway, NJ 08855, U.S.A. } \


\tableofcontents

\vfill \eject

\section{Online material in theory papers in Nature and Science. }\label{}

The past decade has seen the rise in importance of high impact science 
journals like Science, Nature and its spectrum of associated journals,
Nature Physics, Photonics and Materials. 
Funding agencies increasingly look to measure physicists'
performance by the articles they have published in these high impact
journals. 

The established format for papers in these high impact journals tends
to minimize the number of mathematical equations, favoring a more
conceptual and richly colored figure-based representation of key
results.  This format is ideal for experimental papers, but puts
theoretical papers that rely on the language of mathematics at a
disadvantage.  We believe that the supporting online materials that
accompany Science and Nature articles can provide a new format that
can help redress this balance. More mathematical theory papers that are
submitted to these journals can now be written with the main
conceptual results in the body of the paper, accompanied by key
computations and appendix material online.  Of course, such material
can, in the course of time, be groomed for publication in a longer
article, but in the mean time, this provides a mechanism for key
theory papers to be published in high impact journals. 

The online material presented here provides the background material
for reproducing our results.  We have given an introduction to
symplectic spins, and a full derivation of the the application of
symplectic N to the new heavy electron superconductors $\pu $ and $\np $.
We have also included a brief section describing the test-bed
application of this same method to frustrated magnetism. 

\section{\bf   Symplectic spins}

\subsection{N-dimensional Symplectic Pauli Matrices}\label{}

Symplectic spin operators form a subset of the generators of the $SU
(N)$ group.  
To determine their general form, we simply project
out the component $S$ of the $SU (N)$ spin generators which reverses under
time-reversal, i.e the components $S$ for which 
$\hat \epsilon S^{T} \hat \epsilon^{T} = -S$. 
For even $N$, the fundamental  $SU (N)$ spin generators can be written
\begin{equation}\label{}
[{\cal  T}^{pq}]_{\alpha \beta }= \delta^{p}_{\alpha
}\delta^{q}_{\beta }- \frac{1}{N}\delta^{pq}\delta_{\alpha \beta }.
\end{equation}
Here, all indices range over $ [\pm 1, \pm k]$, (excluding zero) 
where $N=2k$ is even.
The general symplectic spin operator is obtained by
subtracting the time-reversed  $SU (N)$
generator $\hat \epsilon{\cal T}^{T}\hat \epsilon^{T}$ from $\cal T$, 
$S^{pq}= {\cal
T}^{pq}-\hat \epsilon[{ \cal T}^{pq}]^{T}\hat \epsilon^{T}$. 
Putting $[\hat \epsilon]^{\alpha }_{ \beta }= \tilde{\alpha }
\delta^{\alpha }_{-\beta }$, where $\tilde{\alpha }= {\rm sgn} (\alpha
)$, then 
\begin{equation}\label{explicit}
[S^{pq}]_{\alpha \beta } 
=\delta^{p}_{\alpha }\delta^{q}_{\beta }  - \epsilon^{p}_{\beta }\epsilon^{q}_{\alpha }= \delta^{p}_{\alpha }\delta^{q}_{\beta } 
- 
\tilde{\alpha } \tilde{\beta }\delta^{p}_{-\beta }\delta^{q}_{-\alpha } .
\end{equation}
This traceless matrix satisfies $S^{pq}= - \hat \epsilon [S^{pq}]^{T}\hat
\epsilon^{T}$, or $S^{pq}_{\alpha \beta }= -\tilde{\alpha }\tilde{\beta
}S^{pq}_{-\beta -\alpha }$. Since $S^{pq}=-\tilde{p}\tilde{q}S^{{-q \ -p}}$,
we can choose a set of $\frac{N}{2}(N+1)$ independent generators by
restricting $p+q\geq 0$. 
As in the case of $SU (N)$ matrices, Hermitian generators can be
obtained by either symmetrizing, or antisymmetrizing $S^{pq}$ on $p$
and $q$. The resulting matrices form a set of $N$ dimensional
symplectic Pauli matrices, 
\begin{equation}\label{hermitian}
\sigma_{N} ^{a}\in \left\{ {\textstyle
\frac{1
}{{\cal N}_{pq}} }(S^{pq}+S^{qp}), 
{\textstyle \frac{-i
}{{\cal N}_{pq}}} ( S^{pq}-S^{qp})\right\}, \qquad (p\geq\vert q\vert )
\end{equation}
where 
\begin{equation}\label{}
{\cal N
}_{pq}=\left\{\begin{array}{cc}
  \sqrt{2} & (\vert p\vert \neq \vert q\vert ),\cr
 2 & (\vert p\vert = \vert q\vert ),
\end{array}
\right.
\end{equation}
normalizes ${\rm Tr}[\sigma_{N}^{a}\sigma_{N} ^{b} ] =
2\delta_{ab}$ in the same way as Pauli matrices. 
The $D_{N}=\frac{N}{2} (N+1)$ component ``vector'' of
matrices ${\bsig }_{N}$, where $[{\bsig }_{N}]^{a}=
{\sigma^{a}_{N}}$ ($a=1,2 \dots D_{N}$) plays the role of Pauli matrices for $SP (N)$.
As an example, consider $N=4$ where the spinor $\psi $ and 
spin-flip matrix $\hat
\epsilon$ take the form
\begin{equation}\label{}
\psi = \left(\begin{array}{l}
\psi_{1}\cr
\psi_{-1}\cr
\psi_{2}\cr
\psi_{-2}
\end{array} \right), \qquad \qquad \hat \epsilon=
\left\{\begin{array}{cccc}
0 & 1 & 0& 0\cr
-1 & 0 & 0& 0\cr
0 & 0 & 0& 1\cr
0 & 0 & -1& 0
\end{array} \right\}= \left\{\begin{array}{ll}
i \underline{\sigma}_{2} & 0 \cr
0 & i \underline{\sigma}_{2} 
\end{array} \right\}
\end{equation}
In this case, there are $10$ symplectic matrices
\begin{equation}\label{}
\bsig_{N}=\left\{\left(
\begin{array}{ll}
 \vec\sigma  & 0 \\
 0 & 0
\end{array}
\right),{\textstyle\frac{1}{\sqrt{2}}}
\left(
\begin{array}{ll}
 0 & {\vec\sigma } \\
{\vec\sigma } & 0
\end{array}
\right),\left(
\begin{array}{ll}
 0 & 0 \\
 0 & \vec\sigma 
\end{array}
\right),{\textstyle\frac{1}{\sqrt{2}}}\left(
\begin{array}{ll}
 0 & -i\underline{1} \\
i\underline{1} & \ \ 0
\end{array}
\right)\right\},
\end{equation}
where $\vec{\sigma }= (\sigma^{1},\sigma^{2},\sigma^{3})$ denotes the
three possible choices of Pauli matrix.

\subsection{Dot Product}\label{}

Here we derive the ``dot product'' between two
symplectic spins. 
Any even dimensional matrix can be divided up 
into a symplectic and an antisymplectic part $M
= M_{S}+ M_{A}$, where 
$M_{S}= -\epsilon{M}_{S}^{T}\hat \epsilon^{T}$ and 
$M_{A}= \hat \epsilon{M_{A}}^{T}{\hat \epsilon^{T}}
$. 
The symplectic part is obtained by projection, ${M_{S}}=
{\rm P}M$, where ${\rm P}{M_{A}}=0$ removes the antisymplectic
component.  
Now since 
${M_{A}}+ \hat \epsilon{M_{A}}^{T}{\hat \epsilon^{T}}=0$, it follows that 
\begin{equation}\label{}
{\rm P}M =
\frac{1}{2} (M-\hat \epsilon M^{T}\hat \epsilon^{T} )
\end{equation}
In components
\begin{eqnarray}\label{l}
{\rm P}^{\alpha \beta }_{\gamma\delta}M^{\delta \gamma} &=& \frac{1}{2} 
[M^{\alpha \beta }- {\epsilon }^{\alpha }_{\gamma}M^{\delta \gamma} \epsilon^{\beta }_{\delta  }]\cr
&=& 
\frac{1}{2} 
[\delta^{\alpha }_{\delta }\delta^{\beta }_{\gamma}- { \epsilon }^{\alpha }_{\gamma}
 \epsilon^{\beta }_{\delta }]
M^{\delta \gamma},
\end{eqnarray}
so that 
\begin{equation}\label{projection}
{\rm P}^{\alpha \beta}_{\gamma\delta } = 
\frac{1}{2} 
[\delta^{\alpha }_{\delta }\delta^{\beta }_{\gamma}- { \epsilon }^{\alpha }_{\gamma}
 \epsilon^{\beta }_{\delta }].
\end{equation}
Now, we can always expand $M_{S}=\sum_{a}m_{a}\sigma_{N}^{a}$ in terms of
symplectic Pauli matrices, and with the normalization
${\rm Tr}[\sigma_{N}^{a}\sigma_{N}^{b}]= 2 \delta_{ab}$, $m_{a}=\frac{1}{2}{\rm
Tr}[\sigma_{N}^{a}M]$, so 
 ${\rm P}M =\frac{1}{2}\sum_{a}
{\rm Tr}[ \sigma_{N}^{a}M] \sigma_{N}^{a} $. 
Expanding both sides gives
\begin{equation}\label{l}
{\rm P}^{\alpha \beta }_{\gamma\delta }M^{\delta \gamma}=\frac{1}{2}\sum_{a}[\sigma ^{a}_{\gamma\delta }M^{\delta \gamma}]\sigma ^{a}_{\alpha \beta },
\end{equation}
(where we have temporarily dropped the label $N$ on the $\sigma_{N}$ matrices)
or 
\begin{equation}\label{}
{\rm P}^{\alpha\beta }_{\gamma\delta }=
\frac{1}{2}\sum_{a}\sigma ^{a}_{\alpha \beta }\sigma ^{a}_{\gamma\delta }.
\end{equation}
Inserting (\ref{projection}), we obtain an explicit expression for the
expansion of the dot product between symplectic matrices
\begin{eqnarray}\label{complete}
\sum_{a} (\sigma_{N}^{a})_{\alpha \beta } (\sigma_{N}^{a})_{\gamma \delta }=
[\delta^{\alpha }_{\delta }\delta^{\beta }_{\gamma}- { \epsilon }^{\alpha }_{\gamma}
 \epsilon^{\beta }_{\delta }].
\end{eqnarray}
When used to decouple interactions, the first term leads to
particle-hole exchange terms, while the second term introduces
pairing. This same completeness result is also obtained by brute-force
expansion using the explicit spin representation (\ref{explicit}), which leads
to $\half \sum_{p,q}S^{pq}_{\alpha \beta }S^{qp}_{\gamma\delta }=
[\delta^{\alpha }_{\delta }\delta^{\beta }_{\gamma}- { \epsilon }^{\alpha }_{\gamma}
 \epsilon^{\beta }_{\delta }]
$.

\subsection{Abrikosov Pseudo-Fermion representation }\label{}

Antisymmetric  representations of symplectic spins are obtainsed using
Abrikosov pseudo-fermions\cite{abrikosov}. An explicit expression for the symplectic spin operator 
is given by
\begin{equation}\label{fspin}
\hat S^{pq}= f\dg_{\alpha }[S^{pq}]_{\alpha \beta }f_{\beta }=
 [f\dg_{p}f_{q}- \tilde{p}\tilde{q}f\dg _{-q}f_{-p} 
].
\end{equation}
where, as before $\tilde{p}= {\rm sgn} (p)$. (Note the use of the
carat over $\hat  S^{pq}$ to delineate the quantum operator from the
matrix $S^{pq}$. )
 The corresponding Hermitian spin operators can be obtained by
symmetrizing and antisymmetrizing on $p$ and $q$, as described in
(\ref{hermitian}), writing $\hat {\bS}=
 f\dg_{\alpha} ({\bsig}_{N})_{\alpha \beta }f_{\beta }
$. 
Using the dot product relation
(\ref{complete} ), we can relate these two forms for the spin operator
via
\begin{eqnarray}\label{l}
\hat {\bS}\cdot ( { \bsig }_{N})_{pq }&=& 
f\dg_{\alpha}f_{\beta } 
(\sigma^{a}_{N})_{\alpha \beta }(\sigma^{a}_{N})_{pq}
\cr
&=&  f\dg_{\alpha}f_{\beta } [\delta^{\alpha }_{q }\delta^{\beta }_{p}- { \epsilon }^{\alpha }_{p}
 \epsilon^{\beta }_{q}].
\cr 
&=&f\dg_{q}f_{p}-
\tilde{p}\tilde{q}f\dg_{-p}f_{-q}= \hat {S}^{qp}
\end{eqnarray}
so that 
\begin{equation}\label{relation}
\hat S^{pq}= {\bS}\cdot (\bsig^{T}_{N})_{pq} .
\end{equation}

\subsubsection{SU (2) gauge symmetry}\label{}

To examine the properties of these symplectic spins, 
it is convenient to introduce the pairing operators
\begin{eqnarray}\label{pair}           
{\Psi }\dg &=& 
\sum_{\alpha >0}f\dg_{\alpha }f\dg_{-\alpha }
= {\textstyle{\frac{1}{2}}}
\sum_{\alpha \in [\pm 1,\pm k]}
 \tilde{\alpha }f\dg_{\alpha }f\dg_{-\alpha }
\cr
{  \Psi }&=& 
\sum_{\alpha >0}f_{-\alpha }f_{\alpha }=
{\textstyle{\frac{1}{2}}}\sum_{\alpha \in [\pm 1,\pm k]
}
\tilde{\alpha }f_{-\alpha }f_{\alpha }
\end{eqnarray}
that 
describe the creation and annihilation of time-reverse pairs of
f-electrons.
It is also convenient to introduce the isospin vector 
$\vec{\Psi }= (\Psi_{1},\Psi_{2},\Psi_{3})$,
where $\Psi_{1}= (\Psi\dg + \Psi )$, $\Psi_{2}= -i (\Psi \dg -\Psi )$
and $ \Psi_{3}= \sum_{\alpha >0} ( f\dg_{\alpha }f_{\alpha }- f_{-\alpha
}f\dg_{-\alpha })= n_{f}- \Nhalf$,  which satisfy an $SU (2)$ algebra
$[\hat \Psi_{a},\hat  \Psi_{b}]= 2 i \epsilon_{abc}\hat \Psi _{c}$.
The inversion of spin under time reversal ensures that the pair
creation operators $\Psi\dg   $ creates a spin-singlet of fermions, 
i.e $\Psi \dg $ commutes with the spin operator. 
To see this explicitly,
note that the commutator with the creation operator 
\begin{equation}\label{1stcomm}
[\hat S^{pq},f\dg _{\alpha }]= 
f\dg _{\beta}[S^{pq}]_{\beta\alpha },
\end{equation}
where we have used an index summation convention over $\beta \in [\pm 1,\pm k]$.
The commutator with the corresponding  time-reversed fermion
yields the reversed spin 
\footnote{To see this, take the transpose of (\ref{1stcomm}) to obtain
$[\hat S^{pq},f\dg_{\alpha }]= (S^{pq})^{T}_{\alpha \beta}f\dg_{\beta}$. 
Now, since $\hat \epsilon (S^{pq})^{T}\epsilon^{T}= - S^{pq}$, it
follows that $(S^{pq})^{T}= - \epsilon^{T} (S^{pq})\epsilon$, thus
(\ref{1stcomm}) 
becomes $[ \hat S^{pq},f\dg_{\alpha }] = - (\hat \epsilon^{T}
S^{pq}\hat \epsilon)_{\alpha \beta } f\dg_{\beta }$. Multiplying both
sides by $\hat \epsilon$ and using $\hat \epsilon\cdot\hat\epsilon^T=1
$, we get $[\hat S^{pq}, (\hat \epsilon f)_{\alpha }]= -
S^{pq}_{\alpha \beta } (\epsilon f\dg )_{\beta }$. 
 }
\[
[\hat S^{pq},\tilde\alpha f\dg _{-\alpha }]
= -[S^{pq}]_{\alpha \beta } (\tilde{\beta
}f\dg _{-\beta}),
\]
so that 
\begin{eqnarray}\label{l}
[\hat S^{pq},\tilde{\alpha }f\dg _{\alpha
}f\dg _{-\alpha }]
=f\dg_{\alpha }\left[S^{pq}_{\alpha \beta }- S^{pq}_{\alpha \beta }
\right]\tilde{\beta }f_{-\beta }= 0.
\end{eqnarray}
Thus the pair operator $[\Psi \dg, \hat S^{pq}]=0 $.  The importance
of this relation works both ways: the pair is invariant under $SP (N)$
spin rotations generated , while the $SP (N)$ spin generator 
is invariant under the particle-hole rotations generated by $\Psi \dg $.

The odd parity of spin operators under time reversal thus ensures that they
not only commute with the particle number $n_{f}$, 
they also commute with the pair operators 
\begin{equation}\label{}
[n_{f},\hat S^{pq}]=[\Psi ,\ \hat S^{pq}] = [ \Psi \dg ,\ \hat S^{pq}] =0.
\end{equation}
These identities imply that the spin is invariant under
continuous  Boguilubov transformations  
\begin{equation}\label{}
f_{\alpha}\longrightarrow uf_{\alpha }+ v \  {\rm  sgn} (\alpha )f\dg
_{-\alpha }\qquad \qquad \underline{\hbox{SU (2) symmetry}}
\end{equation}
where $\vert u\vert^{2}+ \vert v\vert^{2}=1$.  In this way, for fermionic spins, time reversal symmetry gives rise
to an $SU (2)$ gauge symmetry of symplectic spins. This symmetry was first
discovered for spin $1/2$ by Affleck et al\cite{affleck}. The above
reasoning extends their work, and identifies the $SU (2)$ gauge
symmetry as gauge symmetry that survives for the (fermionic)
generators of $SP (N)$ for all N.

By contrast, had we carried out the same calculation using   dipole
operators, $\hat {\cal  P}^{pq}=f\dg_{p}f_{q}+\tilde{p}\tilde{q}f\dg_{-q}f_{-p}$,  which
do not invert under time reversal, we would find that the commutator of the
operator with the time-reversed fermion does not change sign, so that
\begin{eqnarray}\label{l}
[\hat {\cal P}^{pq},\tilde{\alpha }f\dg _{\alpha
}f\dg _{-\alpha }]
=f\dg_{\alpha }\left[{\cal P}^{pq}_{\alpha \beta }{\bf +}{\cal P}^{pq}_{\alpha \beta }
\right]\tilde{\beta }f_{-\beta }\ne 0.
\end{eqnarray}
This means that the pair operator $\Psi $ is not a singlet under the
action of the dipole operators. It also means that the dipole
operators are not invariant under particle-hole transformations. 
Indeed, if  the 
Hamiltonian contains {\sl any }dipole spin operators, 
it no longer commutes with the singlet pair operator $\Psi $, reducing
the gauge symmetry back to a $U (1)$ gauge symmetry.

To fully expose the $SU (2)$  gauge-invariance, it is useful to introduce 
the Nambu spinor
\begin{equation}\label{nambus}
\tilde{f}_{\alpha}=\left(
\begin{matrix}
f_{\alpha }\cr
\tilde{\alpha }f\dg_{-\alpha}
\end{matrix}
\right),\qquad \qquad 
\tilde{f}\dg _{\alpha}= (f\dg_{\alpha}, \tilde{\alpha}f_{-\alpha}),
\end{equation}
By direct expansion, the dot product of these two spinors is
the symplectic spin operator:
\begin{equation}\label{nambus2}
\tilde{f}\dg_{p}\cdot \tilde{f}_{q}= f\dg_{p}f_{q}+
\tilde{p}\tilde{q}f_{-p}f\dg_{-q} = \hat S^{pq}+ \delta_{pq}
\end{equation}
i.e.
\begin{equation}\label{}
\hat S^{pq}= \tilde{f}\dg _{p}\cdot \tilde{f}_{q}- \delta_{pq}
\end{equation}
Under the SU (2) gauge transformation, $\tilde{f}_{q}\rightarrow
g\tilde{f}_{q}$  and $\tilde{f}_{p}\dg
\rightarrow \tilde{f}\dg_{p}g\dg $ 
where $g = \left({u \atop v^{*}} {v\atop -v^{*}} \right)$ 
is an $SU (2)$ matrix, so that $\tilde{f}\dg_{p}\cdot g\dg g \cdot
\tilde{f}_{q}=
\tilde{f}\dg_{p}\cdot
\tilde{f}_{q}$
is explicitly $SU (2)$ gauge invariant.


\subsubsection{Constraint}\label{}

To faithfully represent the spin as an irreducible
representation, we need to fix the value of its 
Casimir $ \vec{S}^{2}$. 
If we compute the
Casimir 
using the completeness
relation  (\ref{complete}), we obtain
\begin{eqnarray}\label{cascalc}
\hat {\bS}^{2}&=& \sum_{a\in g} 
(
f\dg_{\alpha } (\sigma ^{a}_N)_{\alpha \beta }
f_{\beta })
(f\dg_{\gamma}(\sigma ^{a}_N)_{\gamma\delta }
f_{\delta })\cr
&=& 
(
f\dg_{\alpha }f_{\beta })
(f\dg_{\gamma}f_{\delta })
[ \delta_{\alpha \delta}\delta_{\beta \gamma} + {\epsilon}_{\alpha
\gamma}
{\epsilon}_{\delta \beta  }].
\end{eqnarray}
This expression can also be obtained by directly expanding the
unconstrained sum
$\frac{1}{2}\sum_{p,q}\hat S^{pq}\hat S^{qp}$. If we normal order the fermion
operators in the second term, 
we obtain 
\begin{eqnarray}\label{l}
\hat {\bS}^{\ 2}
&=& 
(
f\dg_{\alpha }f_{\beta})
(f\dg_{\beta}f_{\alpha}) 
+ \tilde{\alpha }
\tilde{\beta }
(f\dg_{\alpha }f_{-\beta})
(f\dg_{-\alpha }f_{\beta})
\cr
&=&
n_{f} (N+2-n_{f})
- 
\sum_{\alpha ,\beta }(\tilde{\alpha }f\dg_{\alpha }f\dg_{-\alpha }) (\tilde{\beta }f_{-\beta}
f_{\beta}),
\cr
&=&
n_{f} (N+2-n_{f})
- 4 \Psi \dg \Psi 
.
\end{eqnarray}
where $n_{f}= \sum_{\alpha}f\dg_{\alpha}f_{\alpha}$ is the number of
fermions and we have 
introduced the pairing terms defined in (\ref{pair}).
then 
$\Psi = \frac{1}{2}(\Psi_{1}- i \Psi_{2})$, $\Psi\dg  =
\frac{1}{2}(\Psi_{1}+ i \Psi_{2})$ and  $n_{f}=\Psi_{3}+\Nhalf $, so
that 
\begin{eqnarray}\label{l}
\hat {\bS}^{\ 2}
&=& \frac{N}{2} (\frac{N}{2}+2) + 2{\Psi}_{3} -
({\Psi }_{3})^{2}- 
\overbrace{4{ \Psi }\dg {\Psi}}^{{\Psi}_{1}^{2}+{\Psi}_{2}^{2}-i
[{\Psi}_{1},{\Psi}_{2}]}\cr\cr
&=& \frac{N}{2} (\frac{N}{2}+2) - \vec{\Psi}^{2},
\end{eqnarray}
since $[\Psi_{1},\Psi_{2}]= 2 i \Psi_{3}$.
Alternatively, 
\begin{equation}\label{}
{\textstyle \frac{1}{4}} ({\bS}^{2}+ \vec{{\Psi} })^{2}= j (j+1), \qquad \qquad (j = N/4),
\end{equation}
where, since  $N$ is any even number, $j$ is an integer or half-integer.
This useful identity generalizes the well-known property of
conventional spin-$1/2$ fermions, 
expressing the fact that the sum of spin and charge fluctuations are fixed.
In particular, when the isospin is zero $\vec{\Psi }=0$, the
magnitude of the spin is maximized, $\frac{1}{4}{\bS}^{2}=j (j+1)$.
We adopt the spin maximizing constraint 
$\vec{{\Psi} }=0$ 
in all of our calculations. This constraint imposes 
three conditions:
\begin{eqnarray}\label{l}
{\Psi}_{3}\vert \psi \rangle &=& (n_{f}-N/2)\vert \psi \rangle =0,\cr
{\Psi}\dg \vert \psi \rangle & =& \sum_{\alpha>0}f\dg_{\alpha }f\dg_{-\alpha }\vert \psi \rangle  =0,\cr
{\Psi}\vert \psi \rangle & =& \sum_{\alpha>0}f_{\alpha }f_{-\alpha }\vert \psi \rangle  =0.
\end{eqnarray}
The first constraint implies that the state is half-filled, with
$n_{f}=N/2$.  The second and third terms express the fact that to obtain
irreducible representations of the $SP (N)$ group, we must 
project out all singlet pairs from the state
$\vert \psi \rangle $. These additional constraints become
particularly important when we come to examine heavy electron
superconductivity, for they impose the fact that there can be {\sl no
s-wave pairing amongst the heavy electrons}.
In a path integral approach, we impose the above constraints through
the following term in the action
\begin{eqnarray}\label{l}
H_{C}
& =&
 W^{(-)}f\dg_{\alpha }f\dg_{-\alpha } 
+ W^{3} (n_{f}-N/2)
+ W^{(+)}f_{-\alpha }f_{\alpha}\cr
&=& 
\tilde{f}\dg _{\alpha } 
(
{\bf{W}}\cdot 
{\mbox {\boldmath $\tau$}}
)
\tilde{f}_{\alpha },
\end{eqnarray}
where $ {\bf W}
= (W_{1},W_{2},W_{3})$
is a vector boson field  that couples to the isospin ${\mbox
{\boldmath $\tau$}}$ of the f-spinor
and $W^{(\pm)}=W_{1}\pm iW_{2}$. 

\noindent \underline{Aside:}
The three component vector boson
that imposes the neutrality on the f-spins bears close resemblance to
the $W$-boson in the electro-weak theory of Weinberg and Salam.
Indeed,
the appearance of charged heavy electrons from neutral spins can be
closely likened to the Higg's effect that occurs in electro-weak theory.
If we combine the constraint field on the f-electrons with the
potential field acting on conduction electrons, we get
\[
H_{C}= H_{C}
 =
 W^{(-)}f\dg_{\alpha }f\dg_{-\alpha } \raisebox{-0.15in}{
\mbox{
\fbox{
\rule{0pt}{20pt}\parbox{0.7truein}
{\vskip -0.3in\hskip -0.9in\begin{eqnarray}
&+& W^{3} n_{f}\cr
&- &e \Phi n_{c}\nonumber
\end{eqnarray}\vskip -0.1in}
}}}\ 
+ W^{(+)}f_{-\alpha }f_{\alpha}
\]
In the symplectic 
$N$ description of the Kondo effect, the development of a
hybridization in the primary screening channel 
corresponds to a Higg's effect, which leads to the combination
$Z=\frac{1}{2} (W^{3}+e\Phi )$ becoming massive, forming a high energy
plasmon and $\gamma= \frac{1}{2} (W^{3}-e\Phi )$ forming the new photon
that couples equally to $f$ and $c$ conduction electrons, giving the
f-electrons charge.

\subsection{Schwinger Boson representation}

Symmetric representations of the $SP (N)$ group, useful for 
magnetism applications, are obtained in a similar way to the fermionic
representation, using Schwinger bosons\cite{schwinger}, $S^{a}= b\dg_{\alpha} (\sigma_{N}^{a})_{\alpha \beta }b_{\beta
}$, or using (\ref{explicit}), we can write  the more explicit form 
\begin{equation}\label{}
\hat  S^{pq}= b\dg_{\alpha}S^{pq}_{\alpha \beta }b_{\beta }= 
[b\dg_{p}b_{q}- \tilde{p}\tilde{q}b\dg_{-q}b_{-p}].
\end{equation}
However, the bosonic constrain is simpler, as we now show.

By using  the dot-product relationship
 (\ref{complete}) we obtain
\begin{eqnarray}\label{l}
\hat \bS ^{2}&=& \sum_{a\in g} 
(
b\dg_{\alpha } (\sigma ^{a}_{N})_{\alpha \beta }b_{\beta })
(b\dg_{\gamma} (\sigma ^{a}_{N})_{\gamma\delta }b_{\delta })\cr
&=& 
(
b\dg_{\alpha }b_{\beta })
(b\dg_{\gamma}b_{\delta })
[ \delta_{\alpha \delta}\delta_{\beta \gamma} +\epsilon_{\alpha
\gamma}
\epsilon_{\delta \beta  }]\cr
&=& 
(
b\dg_{\alpha }b_{\beta})
(b\dg_{\beta}b_{\alpha}) 
+ \tilde{\alpha }
\tilde{\beta }
(b\dg_{\alpha }b_{-\beta})
(b\dg_{-\alpha }b_{\beta})
\cr
&=&
\left[(b\dg_{\alpha }b_{\alpha}b_{\beta}
b\dg_{\beta}) 
- n_{b}  \right]
+
\left[ 
\tilde{\alpha }
\tilde{\beta }
(b\dg_{\alpha }b\dg_{-\alpha }) (b_{-\beta}
b_{\beta})+n_{b}\right],
\end{eqnarray}
where $n_{b}= \sum_{\alpha}b\dg_{\alpha}b_{\alpha}$ is the number of
bosons.
The $\pm n_{b}$ in the two terms cancel one-another, while, 
for Schwinger bosons, the pairing terms inside the second 
term vanish, $b\dg_{\alpha }b\dg_{-\alpha }\tilde{\alpha }=0$, 
so the final result is
\begin{equation}\label{}
{\bS}^{2}= (b\dg_{\alpha }b_{\alpha}b_{\beta}
b\dg_{\beta})  = 
n_{b} (n_{b}+N)
\end{equation}
The Casimir of the representation is thus set by fixing the number of
bosons. We choose the convention 
\begin{equation}\label{}
n_{b}= N{S},
\end{equation}
where upon
\begin{equation}\label{}
{\bS}^{2}= N^{2}{S} ({S}+1).
\end{equation}

\section{Heavy Fermion Superconductivity}
Here we derive a two-channel
Kondo lattice model for the $\np $ and $\pu $ heavy electron
superconductors and construct the mean-field theory for the symplectic large $N$
limit of our model. We use this to determine 
the critical temperature of the uniform composite pairing instability
in the frame of the symplectic large-$N$ mean field theory. We derive
the Andreev reflection off the composite-paired f-electron and show
how the crystal-field symmetry determines the structure of the gap. 
Following our discussion on the 
mean field theory, we analyze the fluctuation corrections to the NMR relaxation 
rate.

\fg{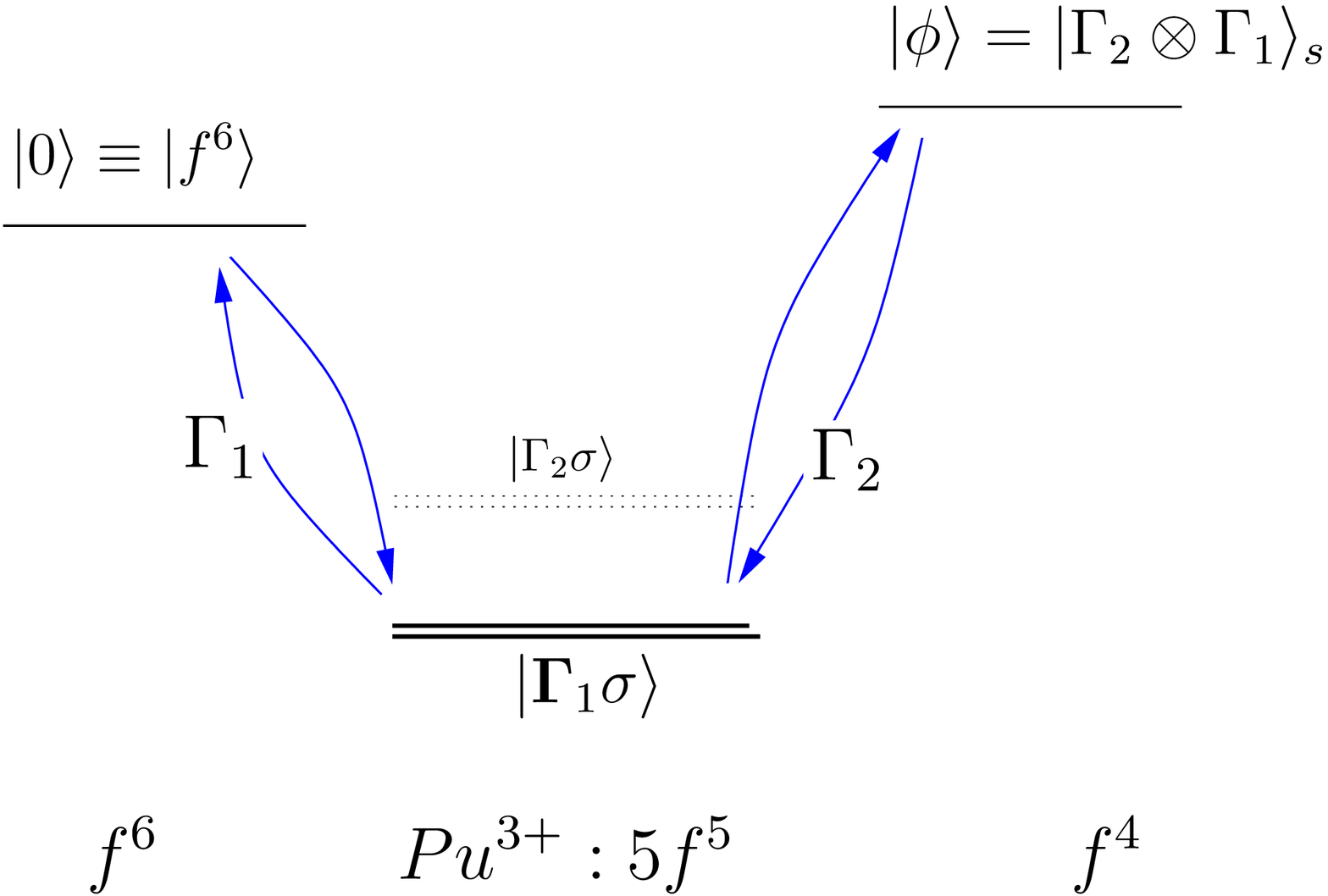}{fig1}{Virtual charge fluctuations of a model
$Pu^{3+}$ ($5f^{5}$) Kramers doublet into singlet states. 
The addition and removal of an f-electron occur in channels
$\Gamma_{1}$ and $\Gamma_{2}$ of different crystal field symmetry.
}

\subsection{Construction of the model.}\label{}

Our two-channel Kondo lattice model for heavy electron superconductivity
assumes that the ground-state of an isolated magnetic ion is a Kramer's doublet $\vert \Gamma_{1}\sigma
\rangle $
containing an odd number
$n$ of f-electrons (Fig. \ref{fig1}). 
In $\pu$ and $\np$, 
the local moments are built out of 
of $f-$electrons in the $5f$  shell. 
The $Pu^{3+}$ ion in $\pu $ is a single f-hole in a
filled $j=5/2$ atomic shell, forming 
a $\vert  5f^{5}\rangle $ Kramer's doublet with $n=5$. The situation
in $\np$ is more uncertain, the Curie moment extracted from the
magnetic susceptibility is closest to that of a $5f^{3}$ ion with
$n=3$.  

We assume that the dominant spin fluctuations occur via
valence fluctuations into singlet states
\begin{eqnarray}\label{l}
\begin{array}{ccccc}
\vert 0\rangle &\rightleftharpoons&\vert \Gamma_{1}\sigma \rangle
&\rightleftharpoons&\vert \phi \rangle\cr
f^{n+1}&&f^{n}&&f^{n-1}.
\end{array}
\end{eqnarray}
To illustrate the situation, consider $\pu$, where $\vert 0\rangle
\equiv \vert f^{6}\rangle $ is a $j=5/2$ f-shell.
In a tetragonal crystal
environment, the sixfold degenerate $j=5/2$ multiplet of f-electrons
splits into three Kramers doublets: $\{\Gamma_7^{+},\
\Gamma_7^{-},\ \Gamma_6\}$. The $f^{5}$ Kramer's doublet can be
written 
\begin{equation}\label{}
\vert
\Gamma_{1} \sigma\rangle = f\dg_{\Gamma_{1}\sigma }\vert 0\rangle  
\end{equation}
where $f\dg_{\Gamma\sigma }$ creates an f-{\sl hole} in one of these
three crystal field states. 
To form a low-energy $f^{4}$ singlet, the strong Coulomb interaction
between f-electrons forces us to add
a second f-hole in a {\sl different} crystal field channel
$\Gamma_{2}$. We assume that this state has the form 
\begin{equation}\label{}
\vert \phi \rangle \equiv \vert \Gamma_{2} \otimes \Gamma_{1}\rangle_{s} =\frac{1}{\sqrt{2}} \sum_{\sigma= \pm 1
}{\rm sgn} (\sigma)f\dg_{\Gamma_{2}\sigma } f\dg_{\Gamma_{1} \ -\sigma }\vert 0\rangle .
\end{equation}
In practice, there are many other excited states, but these are the
most relevant, because they generate antiferromagnetic Kondo
interactions. 
In a conventional Anderson model, $\Gamma_{2}$ and $\Gamma_{1}$ are the same
channel. Hund's coupling forces $\Gamma_{1}$ and $\Gamma_{2}$ to be
different, and it is this physics that introduces new symmetry
channels into the charge fluctuations. 
The simplified ``atomic'' model that describes this system is then
\begin{equation}\label{}
H_{at}= E_{0}\vert 0\rangle \langle 0\vert 
+ E_{1}
\vert \Gamma \sigma
\rangle \langle \Gamma \sigma \vert  + E_{2}\vert \phi \rangle \langle \phi \vert 
\end{equation}
where $E_{1}< E_{0}, E_{2}$. 

In a
tetragonal environment, the three Kramer's doublets are determined by 
\begin{equation}\label{}
f\dg_{\tilde\Gamma\sigma }=\sum_{m\in [-5/2,5/2]}\langle \tilde\Gamma 
\alpha \vert {\textstyle \frac{5}{2} }m\rangle
f\dg_{m\sigma }, \qquad \qquad (\sigma = \pm )
\end{equation}
\figwidth = 14cm
\fg{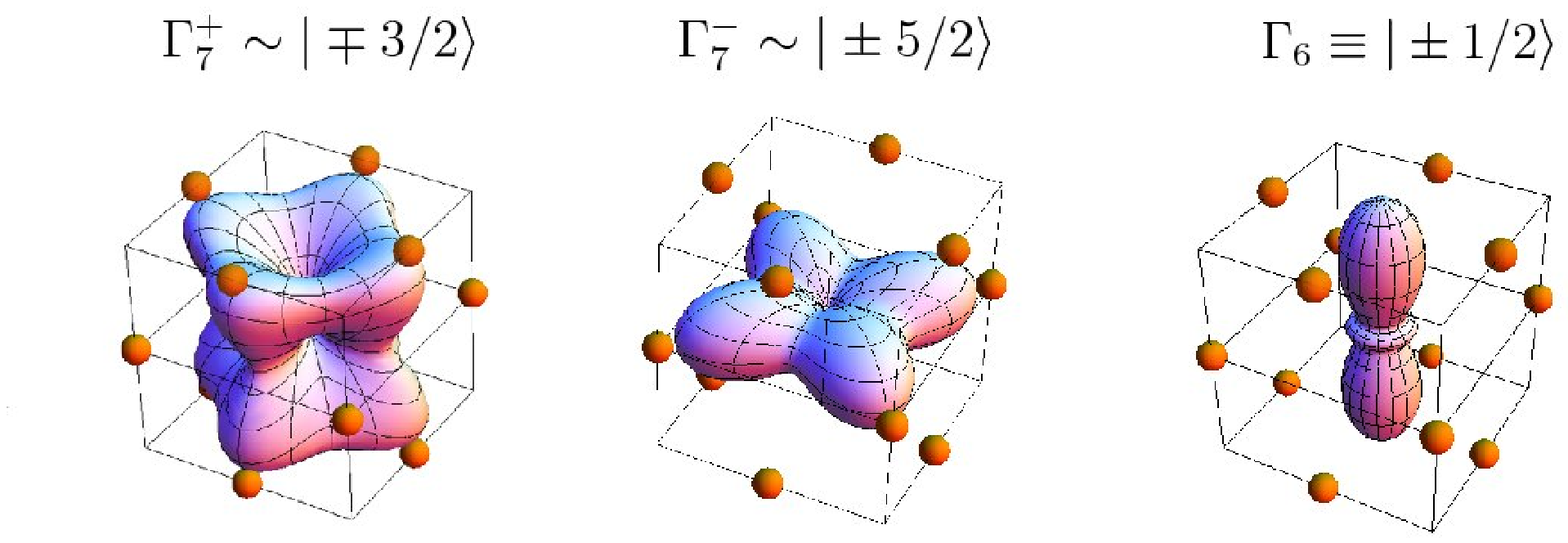}{fig2}{\tiny \sl Showing the three crystal field states
$(\Gamma_{7}^{+},\Gamma_{7}^{-},\Gamma_{6})$  with a mixing angle 
$\beta =
\pi/10$, which maximizes the overlap of the $\Gamma_{7}^{\pm}$ with
the nearby ligand atoms.$\Gamma_{7}^{+}$ overlaps with the eight
out-of-plane ligand atoms, whereas $\Gamma_{7}^{-}$ overlaps strongly
with the four in-plane ligand atoms. 
 $\Gamma_{6}$ is independent of the value of
$\phi $, and is always aligned along the c-axis, with minimum overlap
with nearby ligand atoms. }
where 
\begin{eqnarray}\label{xtalfields}
\begin{array}{lccl}
\Gamma_{6}:&\qquad f\dg_{\Gamma_{6}\pm} &=& f\dg_{\pm 1/2}\cr
\Gamma_{7}^{+}:&\qquad f\dg_{\Gamma_{7}^{+}\pm} &=& \cos \beta  f\dg_{\mp 3/2}+
\sin \beta f\dg _{\pm 5/2}\cr
\Gamma_{7}^{-}:&\qquad f\dg_{\Gamma_{7}^{-}\pm} &=& \sin \beta  f\dg_{\mp 3/2}
-\cos \beta f\dg _{\pm 5/2}.
\end{array}
\end{eqnarray}
Here the mixing angle $\beta $ fine-tunes the spatial anisotropy of the
$\Gamma_{7}^{\pm}$ states (see Fig. \ref{fig2}). Notice how the
crystal mixes $\pm
5/2$ with the $\mp 3/2$ states: this is because the tetragonal
crystalline environment transfers $\pm 4$ units of angular
momentum to the electron. 
A first 
approximation to the crystal field states is obtained by simply
setting $\beta =0$, so that $\Gamma_{7}^{+}\sim \vert \mp 3/2\rangle $ and
$\Gamma_{7}^{-}\sim \vert  \pm 5/2\rangle $.

When this atom is immersed into the
conduction sea, the f-orbitals hybridize with conduction electrons
with the same crystal symmetry. The hybridization Hamiltonian is written
\begin{equation}\label{}
H_{hybr}= \sum_{\sigma }\bigl [
V_{\Gamma_{7}^{+}} \psi \dg_{\Gamma_{7}^{+}\sigma }f_{\Gamma_{7}^{+}\sigma }+ 
V_{\Gamma_{7}^{-}} \psi \dg_{\Gamma_{7}^{-}\sigma }f_{\Gamma_{7}^{-}\sigma }+ 
V_{\Gamma_{6}}\psi \dg_{\Gamma_{6}\sigma }f_{\Gamma_{6}\sigma }
+ ({\rm H.c})\bigr ]
\end{equation}
where $\psi \dg _{\Gamma\sigma }$ creates a conduction electron in a
Wannier state with crystal symmetry $\Gamma$ .  The matrix elements of
this Hamiltonian between the Kramer's doublet and the two excited states
are
\begin{eqnarray}\label{l}
\langle 0\vert H_{hyb}\vert \Gamma \sigma \rangle &=& V_{\Gamma_{1}}\psi \dg_{\Gamma_{1}\sigma }\cr
\langle \phi \vert H_{hyb}\vert \Gamma \sigma \rangle &=& 
V_{\Gamma_{2}}\psi _{\Gamma_{2} -\sigma }\tilde{\sigma },
\end{eqnarray}
where $\tilde{\sigma }= {\rm sgn} (\sigma )$.
Thus the removal of an electron occurs in a {\sl different symmetry 
channel} to the addition of an electron. The projected hybridization
matrix becomes 
\begin{equation}\label{}
H_{hybr}=  \sum_{\sigma=\pm}
\left(V_{\Gamma}\psi \dg_{\Gamma \sigma }\vert 0 \rangle \langle \Gamma
\sigma \vert 
+ \tilde{\sigma }
V_{\Gamma_{2}}
\vert \phi \rangle \langle \Gamma \sigma \vert \psi_{\Gamma_{2}-\sigma }
+ {\rm H.c} \right)
\end{equation}
If we now carry out a Schrieffer Wolff transformation that integrates
out the virtual charge fluctuations into the high-energy singlet
states, where the energy of the absorbed, or emitted conduction
electron is neglected, assuming it lies close to the Fermi energy,
then we obtain
\begin{equation}\label{}
H_{K}= -\sum_{\sigma', \sigma=\pm 1 } 
\left (
J_{1}
\vert \Gamma\sigma'
\rangle \psi \phantom{\dg}_{\Gamma_{1}\sigma'}
\psi \dg _{\Gamma_{1}\sigma } \langle \Gamma_{1}\sigma \vert 
+ J_{2}\ 
\tilde{\sigma }'
\psi\dg_{\Gamma_{2}-\sigma }
\vert \Gamma_{1}\sigma'
\rangle \langle \Gamma_{1}\sigma \vert \psi _{\Gamma_{2}-\sigma' }
\tilde{\sigma }\right),
\end{equation}
where
\begin{equation}\label{}
J_{1}= \frac{(V_{\Gamma_{1}})^{2}}{E_{0}-E_{1}}, \qquad 
J_{2}= \frac{(V_{\Gamma_{2}})^{2}}{E_{2}-E_{1}}, \qquad 
\end{equation}
This Hamiltonian can be re-written in terms of spin operators as follows
\begin{equation}\label{HK}
\hat{H}_K= 
{\textstyle \frac{1}{2}}
\bigl [ J_1 {\mbox{\boldmath $\sigma $}}^{\Gamma_{1}} (0)
+
J_2 {\mbox{\boldmath $\sigma $}}^{\Gamma_{2}} (0)\bigr ]
\cdot{\mathbf S}_{f},
\end{equation}
where we have dropped potential scattering terms and introduced the
notation
\begin{equation}\label{}
{\mathbf S}_{f}=
\sum_{\alpha\beta}\vert {\Gamma_{1}}\alpha\rangle
\bsig_{\alpha\beta}\langle{\Gamma_{1}\beta}\vert .
\end{equation}
for the spin of the Kramer's doublet and
\begin{equation}\label{}
 {\mbox{\boldmath$\sigma $}}^{\Gamma_{1}} (0)= \psi \dg_{\Gamma_{1}\alpha}
 {\mbox{\boldmath $\sigma $}}_{\alpha\beta }\psi _{\Gamma_{1}\beta }, 
\qquad 
 {\mbox{\boldmath $\sigma $}}^{\Gamma_{2}} (0)= \psi \dg_{\Gamma_{2}\alpha}
 {\mbox{\boldmath $\sigma $}}_{\alpha\beta }\psi _{\Gamma_{2} \beta }, 
\end{equation}
for the spin density at the origin in channel $\Gamma_{1} $ and channel
$\Gamma_{2}$.

If we now generalize this derivation to a lattice, the interaction
(\ref{HK}) develops at each site, producing
\beg
\label{HKLM}
\hat{H}=\sum\limits_{{\bk }\sigma}\epsilon_{\bk}c_{\bk \sigma}\dg
c_{\bk \sigma}+
{\textstyle \frac{1}{2}}
\sum_{j}\left[ 
J_1\psi\dg _{\Gamma_{1} j\alpha }
{\mbox{\boldmath $\sigma$}}_{\alpha \beta }\psi_{\Gamma_{1} j\beta }+
J_2\psi\dg _{\Gamma_{2} j\alpha }
{\mbox{\boldmath $\sigma$}}
_{\alpha \beta }\psi_{\Gamma_{2} j\beta }
\right]
\cdot{\mathbf S}_j,
\en
where ${\mathbf S}_{j}$ is the spin operator at site $j$ and
$c\dg_{\bk \sigma }$ creates a conduction electron of momentum $\bk $.
We can relate the Wannier states at site $j$ as follows
\begin{equation}\label{}
	\psi _{\Gamma_{1} j\alpha }= \sum_{k\sigma }[\Phi_{1\bk
	}]_{\alpha \sigma }c_{\bk \sigma }e^{i \bk \cdot
	{\bf{R}}_{j}}, \qquad 	\psi _{\Gamma_{2} j\alpha }= \sum_{k\sigma }[\Phi_{2\bk }]_{\alpha \sigma }c_{\bk \sigma }e^{i \bk \cdot {\bf{R}}_{j}}
\end{equation}
where 
\begin{equation}\label{}
[\Phi_{\Gamma\bk }]_{\alpha \sigma }=\langle k\Gamma \alpha \vert \bk
\sigma \rangle = 
\sum_{m\in [-3,3]}\langle \Gamma\alpha \vert 3m,
\frac{1}{2}\sigma 
\rangle Y^{3}_{m-\sigma } (\hat {\bk} )
\end{equation}
is the form factor of the crystal field state.   The Kondo lattice
Hamiltonian then takes the form 
\begin{equation}\label{}
\hat{H}=
\sum_{\bk \sigma }\epsilon_{\bk}c_{\bk \sigma}\dg
c_{\bk \sigma}+
{\textstyle \frac{1}{2}}
\sum_{\bk,\bk',j}c\dg_{\bk \alpha }\left[
J_{1}\Phi\dg _{1\bk }
\bsig  \Phi_{1\bk' }+
J_{2}\Phi\dg _{2\bk }
\bsig  \Phi_{2\bk '}
 \right]_{\alpha \beta }c_{\bk '\beta }\cdot{\bf S}_{f}e^{i (\bk '-\bk
 )\cdot {\bf R}_{j}}
\end{equation}
For pedagogical purposes,
we work largely with the model in which the matrices $\Phi_{\Gamma\bk }=
\phi_{\Gamma\bk }{\bf 1}$
are taken to be spin-diagonal, giving rise to a simpler form
\begin{equation}\label{}
H= \sum_{\bk \sigma }\epsilon_{\bk}c_{\bk \sigma}\dg
c_{\bk \sigma}+
{\textstyle \frac{1}{2}}
\sum_{\bk,\bk',j}J_{\bk ,\bk '} (c\dg_{\bk \alpha }
\bsig _{\alpha \beta }
 c_{\bk '\beta })\cdot{\bf S}_{f}e^{i (\bk '-\bk
 )\cdot {\bf R}_{j}}
\end{equation}
where
\begin{equation}\label{}
J_{\bk \bk '}= J_{1}\phi_{1\bk }\phi_{2\bk '}+
J_{2}\phi_{2\bk }\phi_{2\bk '}.
\end{equation}
The results obtained using this model are easily generalized to the
spin-anisotropic case by restoring the spin indices to the form factors.
Lastly, we generalize our model from $SU (2)$ to symplectic-$N$
by replacing the Pauli spin  operators $\bsig_{\alpha \beta
}\rightarrow ( {\bsig}_{N})_{\alpha \beta }$, which we write as
\beg\label{HKLMN}
\hat{H}=\sum\limits_{{\bk }\sigma}\epsilon_{\bk}c_{\bk \sigma}\dg
c_{\bk \sigma}+
{\textstyle \frac{1}{N}}
\sum_{j}\left[ 
J_1\psi\dg _{\Gamma_{1} j\alpha }
({\mbox{\boldmath $\sigma$}_{N}})
_{\alpha \beta }\psi_{\Gamma_{1} j\beta }+
J_2\psi\dg _{\Gamma_{2} j\alpha }
({\mbox{\boldmath $\sigma$}_{N}})
_{\alpha \beta }\psi_{\Gamma_{2} j\beta }
\right]
\cdot{\mathbf S}_j,
\en
which in its simpler, spin-isotropic manifestation assumes the form
\begin{equation}\label{}
H= \sum_{\bk \sigma }\epsilon_{\bk}c_{\bk \sigma}\dg
c_{\bk \sigma}+
{\textstyle \frac{1}{N}}
\sum_{\bk,\bk',j}J_{\bk ,\bk '} c\dg_{\bk \alpha }
 c_{\bk '\beta }S^{\beta \alpha }_{f} (j) e^{i (\bk '-\bk
 )\cdot {\bf R}_{j}}
\end{equation}
It is the large $N$ limit of these lattice models  that we have solved
in our paper. 

\subsection{Decoupling scheme and SU (2) symmetry}\label{}

Here we detail our symplectic-$N$ decoupling scheme for the Kondo
lattice. 
To derive the decoupling procedure, let us first focus on the
interaction  at a given site, temporarily suppressing site indices
$j$.  
By applying the the dot-product relation (\ref{complete} ) on the 
the Kondo interaction $H_{K}= \sum_{\Gamma}H_{K\Gamma}$ (\ref{HK}), 
we obtain 
\begin{equation}\label{}
H_{K\Gamma}= \frac{J_{\Gamma}}{N} 
\bsig_{N}
^{\Gamma}\cdot{ \bf S}_{f}=
\frac{J_{\Gamma}}{N}\psi \dg _{\Gamma\alpha}\psi_{\Gamma\beta }f\dg _{\gamma}f_{\delta } (\delta_{\alpha \delta}\delta_{\beta \gamma}+\epsilon_{\alpha \gamma}\epsilon_{\delta \beta } )
\end{equation}
which can be rewritten in the form 
\begin{eqnarray}\label{l}
H_{K\Gamma} 
%
= -\frac{J_{\Gamma}}{N} \sum_{\alpha \beta }
\left[
:
(\psi\dg_{\Gamma\alpha}f_{\alpha })
(f\dg _{\beta }\psi_{\Gamma\beta}):+
(\psi\dg_{\Gamma\alpha}\tilde{\alpha }f\dg _{-\alpha })
(\tilde{\beta }f_{-\beta }\psi_{\Gamma\beta})
\right]
\end{eqnarray}
where the sum over $\alpha $ and $\beta $ runs over the $N=2k$ spin
indices 
$\alpha ,\beta \in [\pm 1, \pm k]$.
(An alternative way to obtain the same result is to write
the interaction as
$H_{K\Gamma}=\frac{J_{\Gamma}}{2N}S_{\Gamma}^{pq}S_{f}^{qp} $
and expand the expression using (\ref{fspin}).)
When we cast this normal ordered interaction inside a path integral,
we can  carry out a Hubbard Stratonovich transformation on both terms,
as follows:
\begin{eqnarray}\label{decoup1}
H_{K\Gamma}\rightarrow \psi \dg_{\Gamma\alpha } (\bar
V_{\Gamma}f_{\alpha }
+ \bar {\Delta}_{\Gamma}f\dg_{-\alpha }\tilde{\alpha }) + {\rm H.c} + N
\left(\frac{\bar V_{\Gamma}V_{\Gamma}+ \bar \Delta_{\Gamma}\Delta_{\Gamma}}{J_{\Gamma}} \right).
\end{eqnarray}
This Hamiltonian has a local $SU (2)$ gauge invariance under the transformations
\begin{equation}\label{localg}
\left(
\begin{matrix}V_{\Gamma}\cr \Delta_{\Gamma}\end{matrix} 
\right) 
\rightarrow g \left(\begin{matrix}V_{\Gamma}\cr \Delta_{\Gamma}
\end{matrix}  \right) , \qquad 
\left(\begin{matrix}f_{\alpha }\cr \tilde{\alpha }f\dg _{-\alpha } \end{matrix} \right) 
\rightarrow g
\left(\begin{matrix}f_{\alpha }\cr \tilde{\alpha }f\dg _{-\alpha } \end{matrix} \right) 
\end{equation}
where $g = \left({u \atop v} {v^{*}\atop -v^{*}} \right)$ is an $SU
(2)$ matrix. 

The exact solution of the 
symplectic large $N$ limit is provided by the saddle point 
where  $V_{\Gamma}$ and $\Delta_{\Gamma}$ acquire constant expectation
values.  At first sight, it might be thought that the mean field
erroneously predicts superconductivity under all circumstances!
However, provided there is only one channel at each site, 
the effect of the pairing field in (\ref{decoup1}) at each site 
can always be absorbed by a gauge transformation in which
\begin{equation}\label{}
\bar V f_{\alpha} + \bar  \Delta \tilde{\alpha }f_{- \alpha }\dg
\rightarrow  \sqrt{\vert V\vert^{2}+\vert \Delta\vert^{2} } f_{\alpha }.
\end{equation}

We now restore the site label $j$ to all variables. 
It is convenient to recast the decoupled Hamiltonian that results in a Nambu
notation, writing 
\begin{equation}\label{decoupled}
H_{\Gamma K}= \sum_{j,\alpha >0} \left[
(\tilde{\psi }\dg_{j\Gamma\alpha
}{\cal V}\dg_{\Gamma j} \tilde{f}_{j\alpha })+ (\tilde{f }\dg _{j\alpha
}){\cal V}_{\Gamma j}\psi_{\Gamma j\alpha }\right]+
\frac{N}{2J_{\Gamma}}
{\rm Tr}[{\cal V}\dg_{\Gamma j}{\cal  V}_{\Gamma j}]
\end{equation}
where
\[
\tilde{f}_{j\alpha }=\left(
\begin{matrix}
f_{j\alpha }\cr
\tilde{\alpha }f\dg_{j -\alpha }
\end{matrix}
\right),\qquad 
\tilde{\psi }_{\Gamma j\alpha }=
\left(\begin{matrix}
\psi_{\Gamma j \alpha }\cr
\tilde{\alpha }\psi \dg_{\Gamma j -\alpha }
\end{matrix} \right)
\]
are the Nambu spinors for the f-electron and conduction electron in
channel $\Gamma$, while
\[
{\cal V}_{\Gamma j}= \cmatrix{V_{\Gamma j}}{\bar \Delta_{\Gamma j} }{
\Delta _{\Gamma j}}{-\bar V_{\Gamma j}}
\]
describes the Hybridization in channel $\Gamma$ at site j. The
summation over $\alpha $ is restricted to positive values to avoid overcounting.

We seek uniform
mean-field solutions, where ${\cal V}_{\Gamma j}$ is constant at each site.
In this situation, its convenient to re-write the Wannier states
and f-states in a momentum state basis, 
\[
\tilde{f}_{j\alpha }= \frac{1}{\sqrt{{\cal  N}_{s}}}
\sum_{\bk }\tilde{f}_{\bk \alpha}e^{i \bk\cdot \bR_{j}},\qquad 
\tilde{\psi }_{\Gamma j\alpha }=
\frac{1}{\sqrt{{\cal N}_{s}}}
\sum_{\bk }\phi_{\Gamma\bk }\tilde{c}_{\bk
\alpha}
e^{i \bk\cdot \bR_{j}}
\]
where ${\cal N}_{s}$ is the number of sites and 
\[
\tilde{f}_{\bk\alpha }=
\left(\begin{matrix}
f_{\bk \alpha }\cr
\tilde{\alpha }f\dg _{-\bk - \alpha }
\end{matrix} \right), \qquad 
\tilde{c}_{\bk\alpha }=
\left(\begin{matrix}
c_{\bk \alpha }\cr
\tilde{\alpha }c\dg _{-\bk - \alpha }
\end{matrix} \right).
\]
Written in momentum space, the mean-field Hamiltonian is then
\begin{eqnarray}\label{l}
H=
\sum_{\bk, \alpha >0 }\left[ 
\epsilon_{\bk } \tilde{c}\dg_{\bk\alpha}
\tau_{3 } \tilde{c}\dg_{\bk\alpha}
+ \tilde{c}\dg _{\bk\alpha}{\cal V}_{\bk }\dg \tilde{f}_{\bk \alpha}
+ \tilde{f}\dg _{\bk\alpha}{\cal V}_{\bk } \tilde{c}_{\bk \alpha}
+ \lambda \tilde{f}_{\bk \alpha}\tau_{3}\tilde{f}_{\bk \alpha}\right]
+ N{\cal N}_s \left( \frac{
{\rm Tr}[{\cal V}\dg_{1}{\cal  V}_{1}]
}{2J_{1}}
+\frac{
{\rm Tr}[{\cal V}\dg_{2}{\cal  V}_{2}]
}{2J_{2}}
\right).
\end{eqnarray}
We can group these terms into a matrix that concisely describes the
mean-field theory as follows
\begin{eqnarray}\label{l}
H=
\sum_{\bk, \alpha >0 } (c\dg_{\bk\alpha}, \tilde{f}\dg _{\bk\alpha})
\begin{pmatrix}
\epsilon_\bk\tau_{3} & {\cal V}_\bk^\dagger \\
{\cal V}_\bk & \lambda\tau_{3}
\end{pmatrix}
\left( \begin{matrix}
\tilde{c}_{\bk \alpha}\cr \tilde{f}_{\bk \alpha}
\end{matrix}\right)
+ N{\cal N}_s \left( \frac{
{\rm Tr}[{\cal V}\dg_{1}{\cal  V}_{1}]
}{2J_{1}}
+\frac{
{\rm Tr}[{\cal V}\dg_{2}{\cal  V}_{2}]
}{2J_{2}}
\right).
\end{eqnarray}
where ${\cal V}_\bk={\cal V}_1\phi_{1\hat{\bk}}+{\cal
V}_2\phi_{2\hat{\bk}}
$. 
This form of the mean-field theory can be elegantly generalized
to include the effects of spin-orbit coupled crystal fields by
restoring the two-dimensional matrix structure to the 
form-factors $\Phi_{\Gamma\bk }$,
\[
{\cal V}_\bk
\longrightarrow {\cal V}_1\Phi_{1\hat{\bk}}+{\cal
V}_2\Phi_{2\hat{\bk}}
\]

\subsection{Mean Field Theory}\label{mftheorysec}

To derive the mean-field theory of the 
uniform composite pair state \cite{CATK}, we must diagonalize 
the mean field Hamiltonian
\beg
H_{\bk}=
\begin{pmatrix}
\epsilon_\bk\tau_{3} & {\cal V}_\bk^\dagger \\
{\cal V}_\bk & \lambda\tau_{3}
\end{pmatrix}
\en
with ${\cal V}_{\bk}={\cal V}_1\Phi_{1{\bk}}+{\cal
V}_2\Phi_{2{\bk}}$. A simplified treatment of the theory is obtained 
by assuming that 
$\Phi_{\Gamma\bk }= \phi_{\Gamma\bk }{\bf 1}$ are spin-diagonal. A
more complete treatment using the general matrix form factors is given
in (\ref{mfso}).
To examine the uniform pairing state we fix the gauge so that
hybridization in channel $1$ is in the particle-hole channel, with
$V_{1}=iv_{1}$, $\Delta_{1}=0$, i.e ${\cal V}_1=i v_{1} $ 
while the hybridization
in channel $2$ is in the Cooper channel, $V_{2}=0$ and 
${\cal V}_2=\Delta _{2}\tau_2$. 
The eigenvalues $\omega_{\bk }$ of $H_{\bk
}$  are determined by 
\[
{\rm  det} (\omega \underline{1} - H_{\bk })=\omega^4-2\alpha_\bk\omega^2+\gamma_\bk^2=0.
\]
where we have introduced the notation:
\begin{eqnarray}\label{alphanot}
&\alpha_\bk=v_{\bk +}^2+\frac{1}{2}\left(\epsilon_\bk^2+\lambda^2\right), \quad
\gamma_\bk^2=(\epsilon_\bk\lambda-v_{\bk -}^2)^2+4 (v_{1\bk}v_{2\bk}c_{\bk })^2,\\
&v_{1\bk }= v_{1}\phi_{1\bk }, \qquad v_{2\bk }= \Delta _{2}\phi_{2\bk }, \qquad v_{\bk \pm}^{2}=v_{1\bk}^2\pm v_{2\bk}^2.
\end{eqnarray}
The quantity $c_{\bk }$ measures the amplitude for singlet Andreev
reflection. This quantity is unity for the 
for the simplified spin-diagonal
model. When the form factors $\Phi_{\Gamma\bk }$ contain off-diagonal
components, the above equations still hold, but with the 
definitions
\[
\phi_{\Gamma \bk }^{2}= \frac{1}{2}{\rm Tr}\left[\Phi \dg_{\Gamma\bk
}\Phi_{\Gamma \bk } \right]
\]
\begin{equation}\label{}
c_{\bk}= 
{\rm Tr} \left[\Phi \dg_{2\bk
}\Phi_{1\bk }+ \Phi\dg _{1\bk }\Phi_{2\bk } \right]/(4 \phi_{1\bk
}\phi_{2\bk })
= \frac{Re {\rm Tr \left[\Phi\dg_{2\bk }\Phi_{1\bk } \right]}}{\sqrt{
{\rm Tr}\left[\Phi \dg_{1\bk}\Phi_{1 \bk } \right]
{\rm Tr}\left[\Phi \dg_{2\bk}\Phi_{2 \bk } \right]
}}
\end{equation}

The eigenvalues of $H_{\bk }$ are given by $\omega=\omega_{\bk
\pm}$ and $\omega=-\omega_{\bk \pm}$, where
\beg
\omega_{\bk\pm}=
\sqrt{\alpha_\bk\pm(\alpha_\bk^2-\gamma_\bk^2)^{1/2}}.
\en
The quantity
\[
\Delta_{\bk }\sim v_{1\bk }v_{2\bk }c_{\bk }
\]
plays the role of the gap in the spectrum. 
Quasiparticle nodes develop 
on the heavy fermi surface defined by $\epsilon_{\bk }= v^{2}_{\bk
-}/\lambda$ in directions where $\Delta_{\bk }=0$. 

The mean field equations are obtained by minimizing the free energy 
\beg
{\cal F}
=-NT\sum\limits_{\bk
,\pm}\log[2\cosh(\beta\omega_{\bk\pm}/2)]+ N{\cal N}_{s}\sum_{\Gamma= 1,2}\frac{v_\Gamma^2}{J_\Gamma}
\en
with respect to $\lambda$ and $(v_{\Gamma})^{2}$ ($\Gamma = 1,2$), which yields
\begin{equation}
\begin{split}
&\frac{1}{{\cal N}_s}\sum\limits_{\bk\pm}\frac{\tanh(\omega_{\bk\pm}/{2T})}{2\omega_{\bk\pm}}
\left(\lambda\pm\frac{\lambda\alpha_\bk-\epsilon_\bk(
\epsilon_\bk\lambda-v_{\bk -}^2)}{\sqrt{\alpha_\bk^2-\gamma_\bk^2}}\right)=0,\\
&\frac{1}{{\cal N}_s}\sum\limits_{\bk\pm}\phi_{1{\bk}}^2\frac{\tanh(\omega_{\bk\pm}/{2T})}{2\omega_{\bk\pm}}
\left(2\pm
\frac{
(\epsilon_\bk+\lambda)^2+4(v_{2\bk } s_{\bk }) ^{2}
}
{\sqrt{\alpha_\bk^2-\gamma_\bk^2}}
\right)=\frac{4}{J_1}, \\
&\frac{1}{{\cal N}_s}\sum\limits_{\bk\pm}\phi_{2{\bk}}^2\frac{\tanh(\omega_{\bk\pm}/{2T})}{2\omega_{\bk\pm}}
\left(2\pm\frac{(\epsilon_\bk-\lambda)^2+4(v_{1\bk } s_{\bk }) ^{2}}{\sqrt{\alpha_\bk^2-\gamma_\bk^2}}\right)=\frac{4}{J_2}, 
\end{split}
\label{MFEQNS}
\end{equation}
where we have put $s_{\bk }^{2}=1 - c_{\bk }^{2}$.
In the normal phase either $v_1$ or $v_2$ is nonzero, corresponding to
the development of the Kondo effect in the strongest channel.
Therefore, there are two types of normal phase with two different
Fermi surfaces:
\begin{itemize}
\item  $J_1>J_2$, $v_2=0$ with spectrum
\beg
\omega_{\bk\pm}=\frac{1}{2}\left(\epsilon_\bk+\lambda\pm\sqrt{(\epsilon_\bk-\lambda)^2+4v_{1\bk}^2}\right).\qquad
\en
corresponding to Kondo lattice 
effect in channel 1, and

\item  $J_{2}>J_{1}$, 
 $v_{1}=0$,  with dispersion
\beg
\omega_{\bk\pm}=\frac{1}{2}\left(\epsilon_\bk-\lambda\pm\sqrt{(\epsilon_\bk+\lambda)^2+4v_{2\bk}^2}\right).\qquad
\en
corresponding to a Kondo lattice effect in channel 2.

\end{itemize}
The two normal phases are always
unstable with respect to formation of the composite paired state at
sufficiently low temperature.

To illustrate the method, we carried out a model calculation, in which
the band structure of the conduction electrons 
is derived from the  $3D$ tight binding
model:
\beg
\epsilon_\bk=-2t(\cos k_x+\cos k_y+\cos k_z)-\mu
\en
and $\mu$ is a chemical potential.  
Our choice of the form factors is dictated by the corresponding
crystal structure of the PuCoGa$_5$. 
We take $\Phi_{1\hat{\bk}}=\Phi_{\Gamma_7^{+}\bk} $ for electrons in channel one
and $\Phi_{2\hat{\bk}}=\Phi_{\Gamma_7^{-}\bk }$ for the electrons in
channel two.   

As we lower the temperature, the superconducting instability develops in the weaker channel.  
The critical temperature for the composite pairing instability is determined from equations (\ref{MFEQNS}) by putting $v_2=0^+$. From the third equation with logarithmic accuracy 
we have $\log(T_{K1}/T_c)\simeq1/J_2$ which yields
\beg
T_c\simeq\sqrt{T_{K1}T_{K2}}.
\en
signaling an enhancement of superconductivity for $J_1\simeq J_2$. 

It is instructive to
contrast the phase diagrams of the $SU (N)$
and symplectic large $N$ limits.
In the former, there is a single quantum phase
transition that separates the heavy electron Fermi liquids formed via a
Kondo effect about the strongest channel. In the symplectic treatment,
coherence develops between the channels,  immersing 
the two-channel quantum critical point beneath a
superconducting dome. This is, to our knowledge, the first {\sl controlled}
mean-field theory in which the phenemenon of ``avoided criticality''
gives rise to superconductivity.

\subsection{NMR relaxation rate}\label{}

One of the precursor effects of co-operative interference between the two conduction channels 
appears in the NMR relaxation rate just above the transition temperature. The NMR rate is 
determined by  
\beg
{\cal R}\equiv\frac{1}{T_1T}=-\frac{I^2}{2\pi}\lim\limits_{\omega\to 0}\frac{\text{Im}K_{+-}^R(\omega)}{\omega},
\en
where $I$ is the hyperfine coupling constant, $\omega$ is the NMR 
frequency and $K_{+-}^R(\omega)$ is the Fourier transform of the retarded correlation 
function of the electron spin densities at the nuclear site:
\beg
K_{+-}^R(\omega)=-i\int_{0}^{\infty }
\langle[\hat{S}_{+}({\mathbf 0},t),\hat{S}_{-}({\mathbf 0},0)]\rangle
e^{i\omega t}dt
\en
At the mean field level, $N\to\infty$, the NMR relaxation rate follows a Korringa law. 

Corrections to Korringa relaxation appear in the $1/N^2$ corrections
to the mean field.  To simplify our discussion, we assume that the
Kondo exchange constants are almost degenerate $J_{1}\sim J_{2}$.
In the approach to the superconducting transition, at $T>T_{c}$, in
principle, we need to examine the effects of fluctuations in the
hybridization and pairing amplitudes in both channel one and two.
The anomalous NMR effects we are interested in come from the
fluctuations in the composite pairs, and as such are driven by
the interference between hybridization fluctuations in one channel and pair
fluctuations in the other. 
To simplify our discussion we restrict our
attention to the corrections  induced by the interplay between 
fluctuations in the Kondo hybridization in channel one and pairing
fluctuations in channel two. 
The simplified Hamiltonian for our calculation is then
${H}={H}_{c}+{H}_0+{H}_2$ with 
\beg
\begin{split}
&{H}_0=\sum\limits_\bfk\frac{2}{J_1}{\bar V}_\bfk\hat{V}_\bfk+
\frac{2}{J_2}{\bar \Delta }_\bfk {\Delta }_\bfk,
\\
&{H}_{2}=\sum\limits_{\bfk,\bfq;\sigma}\left(
\phi_{1\hat{\bfk}}f_{\bfk+\bfq\sigma}^\dagger{V}_\bfq
c_{\bfk\sigma}+\phi_{2\hat{\bfk}} \tilde{\sigma }c_{\bfk\sigma}^\dagger{\Delta }_\bfq f_{\bfq-\bfk,-\sigma}^\dagger+\text{H.c.}\right)
\end{split}
\en and  ${H}_c$ describes the conduction electrons. 
We ignore the fluctuations in the constraint fields, which do not
couple to the fluctuations in $\Delta $ and $V$ in the lowest orders
that we are considering. 
We
also neglect the spin-orbit interaction effects by taking the form
factors to be diagonal in spin space. Our goal is to compute the
corrections to the $f$-spin correlator due to the channel
interference, i.e. due to the interactions of between heavy electrons
and fluctuations of slave fields described by $\hat{H}_2$. The
interference corrections to the relaxation rate involve the product
$\phi_{1\hat{\bk}}^{*}\phi_{2\hat{\bk}}$.

The relaxation rate will be governed by the $f$-spin correlations:
\beg K_{ff}({\vec
x};\tau)=-\langle\hat{T}_\tau\hat{f}_\uparrow^\dagger({\vec x},\tau)
\hat{f}_\downarrow({\vec x},\tau) \hat{f}_\downarrow^\dagger(0,0)
\hat{f}_\uparrow(0,0)e^{-\int\limits_0^\beta\hat{H}_2(\tau)d\tau}\rangle_c,
\en 
where $\langle...\rangle_c$ denotes the connected  Green's function obtained
by perturbatively expanding the time-ordered exponential 
in the high-temperature state where the f-electrons and conduction electrons
are decoupled at the mean-field level. 
The leading contribution to the temperature dependence of the relaxation
rate  is governed by the diagram on Fig 1 which describes the 
effect of \emph{intersite} scattering associated with an electron
switching from one symmetry channel to another as it hops from
site to site. To write down an analytic expression for the diagram
(Fig. 1) we employ the Matsubara correlation functions for the $f-$
and $c-$ electrons together with the correlation functions of the
slave fields 
$K_V({\vec r},\tau)=-\langle\hat{T}_\tau\hat{V}({\vec
r},\tau)\hat{V}\dg(0,0)\rangle$ and $K_\Delta({\vec
r},\tau)=-\langle\hat{T}_\tau\hat{\Delta}({\vec
r},\tau)\hat{\Delta}\dg(0,0)\rangle$.  Contrary to the situation of only
one conduction channel, this contribution enhances the screening of
the local spins rather then suppressing it. As in the case of the
composite pair superconductivity, contribution on Fig. 1 originates
from an interference between the electrons in two conduction channels
undergoing a Kondo effect. To compute the relaxation rate we take into
account for the onset of the Kondo screening produces renormalization
of the slave boson propagators.  In the Matsubara representation these
propagators are:
\begin{equation}
{\cal K}_V({\vec p};i\Omega)=\left[\frac{1}{J_1}+\Pi_V(i\Omega)\right]^{-1}, 
\quad {\cal K}_\Delta ({\vec p};i\Omega)=\left[\frac{1}{J_2}+\Pi_\Delta (i\Omega)\right]^{-1},
\end{equation}
where $\Pi_{V,\Delta }(i\Omega)$ are the polarization bubbles associated
with hybridization fluctuations in channel 1 and pair fluctuations in
channel 2, which describe 
the renormalization due to the Kondo scattering. The analytic
expression for the diagram on Fig.1 reads:
\begin{equation}\label{analyticFig1}
\begin{split}
&\frac{T^3}{N^2}\sum\limits_{i\epsilon,i\Omega_V,i\Omega_\Delta }
\sum\limits_{q,p,k_V,k_\Delta }G_f(p-k_V)
G_f(p+q-k_V){\cal K}_V(k_V)\\&\times
G_c(p+q)G_c(p){\cal K}_\Delta (k_\Delta )
G_f(k_\Delta -p-q)G_f(k_\Delta -p),
\end{split}
\end{equation}
where we employed the four vector notation $p=({\vec p},i\omega)$ and 
included the form factors into the definition of the conduction electron propagators. 
The Matsubara frequency summations can be performed by employing the spectral function representation for the correlators in expression (\ref{analyticFig1}). For example,
\begin{equation}
G_{f,c}({\vec p};i\omega)=\int\limits_{-\infty}^{\infty}\frac{d\varepsilon}{\pi}
\frac{\rho_{f,c}({\vec p},\varepsilon)}{i\omega-\varepsilon},
\end{equation}
where $\rho_{f,c}({\vec p},\varepsilon)$ are the corresponding
spectral functions. 
In the high temperature phase, where there is  no expectation value to
the hybridization $V$ or pairing field $\Delta $,
the fluctuation 
propagators are independent of momentum 
$K_{V,\Delta}({\vec p};i\Omega)=K_{V,\Delta}(i\Omega)$. The resulting expression for the
relaxation rate can be
compactly written as follows
\begin{equation}\label{relRate_int}
\frac{1}{T_1T}\simeq\frac{1}{N^2}\int\limits_{-\infty}^{\infty}
W_{fc}(\omega){\cal K}_{\Delta }(\omega){\cal K}_V(-\omega)\frac{d\omega}{2\pi},
\end{equation}
where $W_{fc}(\omega)$ is proportional to $\rho_f(\omega)\rho_c(\omega)$. The integral (\ref{relRate_int})
is dominated by the frequency region near the Fermi surface. Finally, approximating the slave
boson functions with ${\cal K}_{V,\Delta}(\omega)\sim J_{1,2}/\log[(T-i\omega)/T_K]$ \cite{hamann} 
we obtain the following estimate for the relaxation rate
\beg
\frac{1}{T_1T}\sim\frac{1}{N^2}\frac{1}{\log^2(T/T_K)+\pi^2},
\en
Our result for the relaxation rate shows an upturn in $(T_1T)^{-1}$ with decrease
in temperature, in agreement with experimental data of Curro et al. \cite{Curro2005}.
\begin{figure}[tbp]
\includegraphics[width=2.7in]{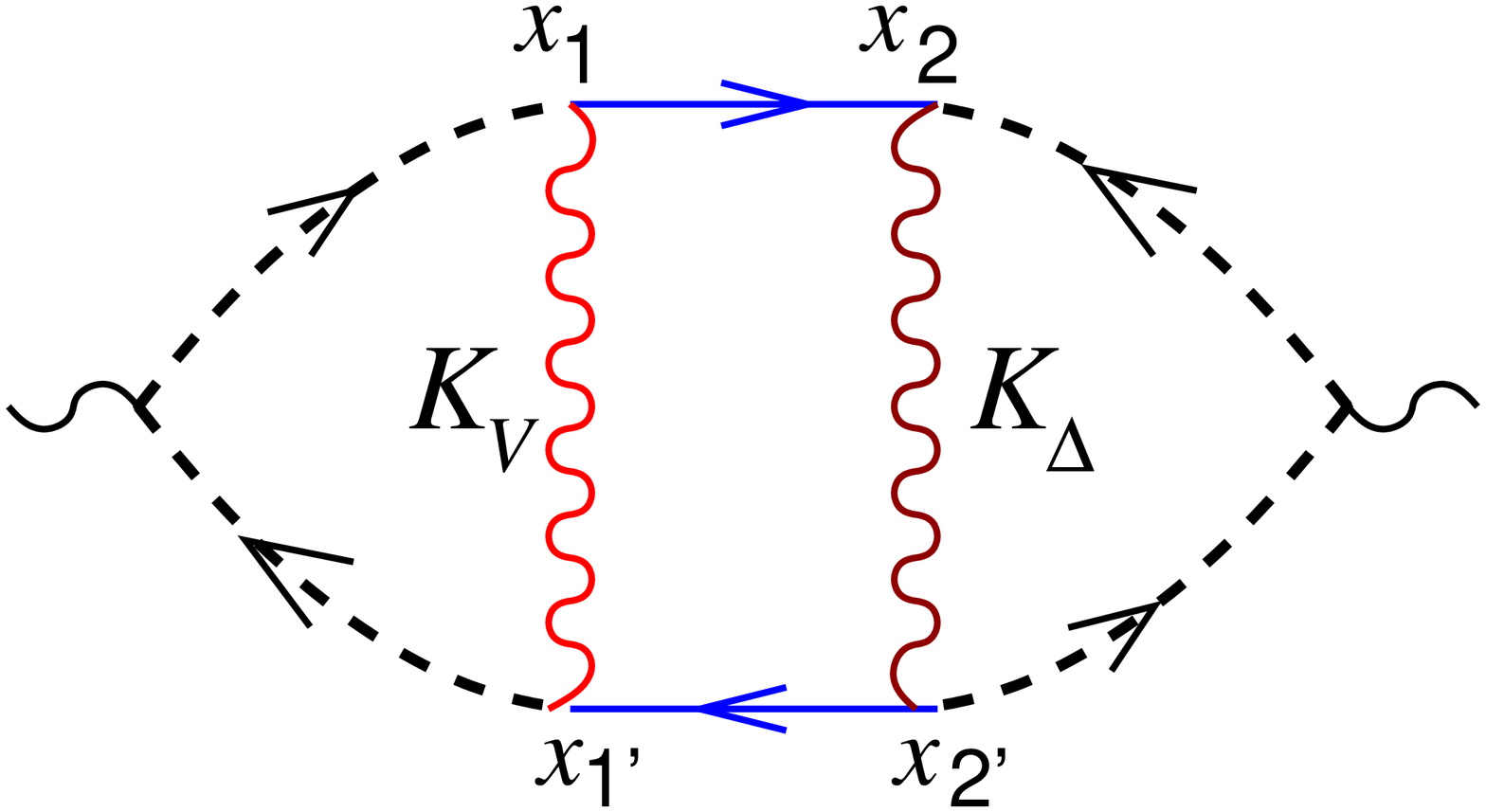}
\caption{Channel interference contribution to the NMR relaxation rate originating from the interaction of nuclear moments with $f$-spins in the normal state. Solid lines are the conduction band propagators, wiggly lines are the propagators $K_V$ and $K_\Delta$ of the slave fields for the channel one and two correspondingly. The dashed lines are the $f$-electron propagators.}
\label{Fig_NMR}
\end{figure}

\subsection{Composite pairing}\label{}

Ostensibly, our mean-field theory is that of a two-band
BCS superconductor, with hybridization processes that pair the heavy
electrons, and Hamiltonian described by
\bea
{\cal H}({\bf k})=
\begin{pmatrix}
\epsilon_\bk\tau_{3} & {\cal V}_\bk^\dagger \\
{\cal V}_\bk & \lambda\tau_{3}
\end{pmatrix}
\eea
However, hidden beneath the hood of theory is the underlying gauge
invariance that maintains the neutrality of the f-spins.
To understand the pairing, we must look not to  the hybridization
pairing terms, which are gauge dependent, but to 
gauge-invariant variables in the theory.
Indeed, it is not possible to say whether the
pairing is channel one, or in channel two. 
In the gauge we have chosen, the 
scattering in channel one is  ``normal'' ${\cal V}_{1}=i v_{1}$ and pairing takes
place in channel two ${\cal V}_{2}= \Delta_{2}\tau_{2}$. 
But suppose 
we make the gauge transformation (\ref{localg}) with
$g_j = -i \tau_2$, then 
\bea
{\cal V}_{\bf k} = i v_{1 {\bf k} }+ v_{2 {\bf k}}\tau_2
&\longrightarrow& i \tau_2 {\cal V}_{\bf k} = 
  v_{1 {\bf k} }\tau_2-i v_{2 {\bf k}},
\cr&&\cr
{\bf W}= \lambda \tau_3 &\longrightarrow&  i \tau_2{\bf W}
(-i \tau_2) = - \lambda \tau_3,
\eea
which transforms the Hamiltonian to one which is now
pairing in channel {\sl one}, and ``normal'' in channel 
two.  The only gauge-invariant statement that we can make, is that 
superconductivity is not a product of one channel or the other, but
instead derives from a coherence between the two channels. 

To see this, we must combine the gauge-dependent order parameters
\begin{equation}\label{}
{\cal V}_{1 }= \cmatrix{V_{1 }}{\bar \Delta_{1} }{
\Delta _{1}}{-\bar V_{1 }}, \qquad\qquad 
{\cal V}_{2 }= \cmatrix{V_{2 }}{\bar \Delta_{2} }{
\Delta _{2}}{-\bar V_{2 }}
\end{equation}
into the gauge-invariant composite
\[
{\cal
V}\dg _{2}{\cal V}_{1} = \cmatrix{
\bar V_{2}V_{1}+\bar\Delta_{2}\Delta_{1}}
{\bar V_{2}\bar \Delta_{1}- \bar V_{1}\bar\Delta_{2}}
{V_{1}\Delta_{2}- V_{2} \Delta_{1}}
{\bar V_{1}V_{2}+\bar\Delta_{1}\Delta_{2}}.
\]
Under an $SU (2)$ gauge transformation, 
${\cal V}_{1}\rightarrow g{\cal V}_{1}$, ${\cal V}_{2}\dg \rightarrow
{\cal V}_{2}\dg g\dg $, so that $
{\cal V}_{2}\dg {\cal V}_{1}
\rightarrow 
{\cal V}_{2}\dg g\dg g{\cal V}_{1} = {\cal V}_{2}\dg {\cal V}_{1}
$ is gauge invariant. 
We shall now show that this matrix is equal to the amplitudes for
composite pairing and hybridization, as follows:
\begin{eqnarray}\label{composite}
{\cal V}\dg _{2}{\cal V}_{1}
 = - \frac{J_{1}J_{2}}{N^{2}}
\zmatrix{
\psi \dg_{1}(
{\mbox {\boldmath $\sigma $}_{N}}
\cdot {\bf S})\psi_{2}
}
{
\psi \dg_{1}
({\mbox {\boldmath $\sigma $}}_{N}\cdot {\bf S})\epsilon \psi \dg_{2}
}
{\psi _{1}\epsilon^{T}({\mbox {\boldmath $\sigma $}}_{N}\cdot {\bf S})\psi_{2}}
{\psi _{1}\epsilon^{T}({\mbox {\boldmath $\sigma $}}_{N}\cdot {\bf S})\epsilon\psi \dg_{2}}
\end{eqnarray}
In the mean-field theory we have developed, ${\cal V}_{1}= iv_{1}$ and
${\cal V}_{2}= \Delta \tau_{2}$, so the composite order parameter is thus
given by
\begin{equation}\label{}
\langle \psi \dg_{1}
({\mbox {\boldmath $\sigma $}}\cdot {\bf S})\epsilon \psi \dg_{2}\rangle 
=-\frac{N^{2}}{ J_{1}J_{2}}
(iv_{1}\Delta_{2})
\end{equation}
Thus it is the combination $v_{1}\Delta_{2}$ that determines the
composite pairing that is ultimately manifested as resonant 
Andreev reflection. (see \ref{andrsc})

To prove identity (\ref{composite}), we use a path integral approach.
Here, it proves useful to employ the following matrix representation
for the conduction and f-fields at each site $j$
\begin{eqnarray}\label{newnotn}
F_{j}= 
\left(\begin{matrix} f_{j}^{T}\cr f_{j}\dg  \hat  \epsilon^{T}
\end{matrix} \right)=
\left(\begin{matrix}
f_{j1}&f_{j-1}& \dots &f_{jk}&f_{j-k}\cr
f\dg_{j-1}&-f\dg _{j1}& \dots &f\dg_{j-k}& -f\dg_{jk}
\end{matrix} \right)_{j}\cr
\Psi _{\Gamma j}=
\left(\begin{matrix} \psi _{\Gamma j}^{T}\cr \psi _{\Gamma j}\dg  \hat  \epsilon^{T}
\end{matrix} \right)=
\left(\begin{matrix}
\psi_{\Gamma j1}&\psi_{\Gamma j -1}& \dots &\psi_{\Gamma
jk}&\psi_{\Gamma j -k}\cr
\psi\dg_{\Gamma j  -1}&-\psi\dg _{\Gamma j  1}& \dots &\psi\dg_{\Gamma
j -k}&
-\psi\dg_{\Gamma  j k}
\end{matrix} \right)
\end{eqnarray}
whose columns are made up the Nambu spinors introduced in (\ref{nambus}).
Using (\ref{nambus2}), we can write 
$
(F_{j}\dg F_{j})_{pq} 
= \hat S^{pq} (j)+ \delta_{pq}$, 
and using  (\ref{relation}), $\hat S^{pq}=
{\bS}\cdot ( {\bsig }^{T}_{N})_{pq}
$, it follows that 
$F_{j}\dg F_{j}
= {\bS}\cdot {\bsig }^{T}_{N}
+ \underline{1}
$.
When we work with a path integral we shall need the normal-ordered
version of this result, 
\begin{equation}\label{important}
:F_{j}\dg F_{j}:
= {\bS}\cdot {\bsig }^{T}_{N}
\end{equation}
.

In the following derivation we temporarily suspend the site index $j$
of clarity. The notation introduced in  (\ref{newnotn}) can
be used to recast the hybridization terms in the interaction
Hamiltonian in a more compact form as follows:
\begin{equation}\label{}
\sum_{\alpha >0}\tilde{\psi }_{\Gamma \alpha }{\cal
V}_{\Gamma}f_{\alpha }
= \frac{1}{2}
\sum_{\alpha}\tilde{\psi }_{\Gamma \alpha }{\cal
V}_{\Gamma}f_{\alpha } = \frac{1}{2}{\rm Tr}\left[\Psi
\dg_{\Gamma}{\cal V}\dg_{\Gamma}F \right] = - \frac{1}{2}
\left[F \Psi \dg_{\Gamma}{\cal V}\dg _{\Gamma} \right] = 
-\frac{1}{2}{\rm Tr}\left[U_{\Gamma}{\cal V}\dg_{\Gamma} \right]
\end{equation}
where $U_{\Gamma}= F\Psi\dg _{\Gamma } $ is a two-dimensional matrix operator.
In terms of
this representation, the decoupled interaction Hamiltonian (\ref{decoupled})
at each site assumes the compact form
\begin{equation}\label{compact}
H_{K} = -\frac{1}{2}\sum_{ \Gamma}  
\left( {\rm Tr}\left[
U_{\Gamma}{\cal V}\dg _{\Gamma}
\right]
 + {\rm H. c.}\right)+\frac{N}{2J_{\Gamma}}
{\rm Tr}[{\cal  V}_{\Gamma }{\cal V}\dg_{\Gamma }
]
\end{equation}
where we have introduced the two dimensional matrices
Now to evaluate the expectation value of the matrix operator ${\cal
M}={\cal V}\dg _{2}{\cal V}_{1}$ we add a source term to $H_{\Gamma
K}$ as follows:
\begin{equation}\label{source}
H_{ K}[\eta ] = H_{K} + {\rm Tr}\left[\eta{\cal
M} + (\rm H. c.)\right].
\end{equation}
By varying $\eta $, we can read off the matrix $\cal M$, 
\[
\frac{\delta H_{\Gamma}}{\delta n_{\beta \alpha }} = ( {\cal M})_{\alpha \beta }
\]
Now we can combine the source term in (\ref{source}) with the final
trace term in (\ref{compact}) to 
rewrite $H_{K}[\eta ]$ in the following form
\[
H_{K}[\eta ]=-\frac{1}{2}\sum_{ \Gamma}  
\left( {\rm Tr}\left[U_{\Gamma }{\cal V}\dg _{\Gamma
}
 \right]
 + {\rm H. c.}\right)+\frac{1}{2}{\rm Tr}\left[{\cal V}_{\Gamma} {\cal
 J}^{-1}_{\Gamma \Gamma'}{\cal V}\dg_{\Gamma'}\right]
\]
where
\[
{\cal J}^{-1}= \zmatrix{\frac{N}{J_{1}}}{2\eta }{2\bar \eta }{\frac{N}{J_{2}}}
\]
If we now carry out the Gaussian integral over the ${\cal
V}_{\Gamma}$, we obtain
\[
H_{K}[\eta ] = - \frac{1}{2}{\rm Tr}\left[U_{\Gamma}{\cal J}_{\Gamma
\Gamma'} U\dg _{\Gamma'} \right]
\]
where to leading order in $\eta $
\[
{\cal J}= \frac{J_{1}J_{2}}{N^{2}}
\zmatrix{\frac{N}{J_{2}}}{-2\eta }{-2\bar \eta }{\frac{N}{J_{1}}}
= \zmatrix{\frac{J_{1}}{N}}{-2\frac{J_{1}J_{2}}{N^{2} }\eta 
}{-2\frac{J_{1}J_{2}}{N^{2} }\bar \eta }{\frac{J_{2}}{N}}
\]
So expanding $H_{K}[\eta ]$ to leading order in $\eta $, we get
\[
H_{K}[\eta ]= - \frac{J_{\Gamma}}{2N}{\rm Tr}\left(U_{\Gamma}U\dg_{\Gamma}
\right) + \frac{J_{1}J_{2}}{N^{2}}{\rm Tr}\left[ \eta U_{2}\dg
U_{1}+{\rm H. c}\right]
\]
Differentiating with respect to $\eta $ then gives
\[
{\cal V}\dg_{2}{\cal V}_{1} \equiv \frac{J_{1}J_{2}}{N^{2}}
U\dg_{2}U_{1}= 
\frac{J_{1}J_{2}}{N^{2}} :\left[ \Psi_{2}F\dg  F \Psi\dg _{1}\right]:.
\]
(where the final transition from Grassman variables $U_{2}\dg U_{1}$ to operators requires the introduction of normal-ordering, denoted by  colons).
Using the identity (\ref{important}) $:F\dg F: = \bsig^{T}_{N}\cdot {\bf S}
 $,
we obtain
\[
{\cal V}\dg_{2}{\cal V}_{1}=
\frac{J_{1}J_{2}}{N^{2}} :\left[ \Psi_{2} (\bsig^{T}_N\cdot {\bf S})
\Psi\dg _{1}\right]: 
= - 
\frac{J_{1}J_{2}}{N^{2}} :\left[ \Psi_{1}^{*} ( \bsig_N\cdot {\bf S}
)
\Psi^{T} _{2}\right]: 
\]
where we have taken the transpose of the expression in the last step. 
Using 
\begin{eqnarray}\label{l}
\Psi _{1}^{*}&=& (\Psi_{1}\dg )^{T}= \left(\dots \begin{matrix}\psi\dg
_{1\alpha }\cr \tilde{ \alpha }\psi_{1 -\alpha }
\end{matrix}\dots  \right) = \left(\begin{matrix} \psi \dg_{1}\cr \psi_{1}^{T}\epsilon^{T}\end{matrix} \right),
\cr\Psi_{2}^{T}&= &\left( \psi_{2}, \epsilon \psi_{2}\dg\right),
\end{eqnarray}
to expand the last term, we obtain
\begin{equation}\label{l}
{\cal V}\dg _{2}{\cal V}_{1}
=- 
\frac{J_{1}J_{2}}{N^{2}} :\left[ \Psi_{1}^{*} (
\bsig_{N}\cdot {\bf S}
)
\Psi^{T} _{2}\right]:  \equiv - \frac{J_{1}J_{2}}{N^{2}}
\zmatrix{
\psi \dg_{1}(
{\bsig_{N}}
\cdot {\bf S})\psi_{2}
}
{
\psi \dg_{1}
({\bsig}_{N}\cdot {\bf S})\epsilon \psi \dg_{2}
}
{\psi _{1}\epsilon^{T}(\bsig _{N}\cdot {\bf S})\psi_{2}}
{\psi _{1}\epsilon^{T}({\bsig}_{N}\cdot {\bf S})\epsilon\psi \dg_{2}}
\end{equation}

\subsection{Resonant Andreev scattering}\label{andrsc}

Composite pairing manifests itself through the development of an Andreev
reflection component to the resonant scattering off magnetic
impurities. We can capture this scattering in the mean-field theory by
integrating out the f-electrons. This leads to a conduction electron
Green's function of the form 
\bea
{ G}(\kappa)^{-1}
= 
\omega - \epsilon_{\bf k} \tau_3 - \Sigma(\kappa),\label{conduction}
\eea
where $\kappa\equiv({\bf k}, \omega)$ 
and 
\bea
\Sigma(\kappa)= 
{\cal V}\dg_{\bf k}
( \omega -  \lambda  \tau_3 )^{-1}{\cal V}_{\bf k}
\eea
describes the resonant scattering off the quenched
local moments.  The hybridization matrices are written 
\begin{equation}\label{}
{\cal V}_{\bk }= {\cal V}_{1}{\bPhi}_{1\bk}+ {\cal V}_{2}{\bPhi}_{2\bk }=
iv_{1}\bPhi_{1\bk }+ \Delta 
\bPhi_{2\bk }\tau_{2}
\end{equation}
In our earlier discussions, the quantities $\Phi_{\Gamma\bk }$ were
assumed to be spin-diagonal.  We now restore their two-dimensional
matrix character, adding a carat to the symbol to denote its matrix
character. 
These matrices act identically on particle and hole states
\footnote{Under a particle-hole transformation, 
$\Phi_{\Gamma \bk }\rarrow\left( \epsilon \ [\Phi\dg _{\Gamma -\bk
}]^{T}\epsilon^{T}\right)$, which corresponds to the time-reversed
form-factor. However, since $\left( \epsilon \ [\Phi\dg _{\Gamma -\bk
}]^{T}\epsilon^{T}\right) 
= \Phi_{\Gamma\bk }$ is time-reverse invariant so the form factor is
invariant under particle-hole transformations, and acts equally on
particles and holes. }
, commuting
with the isospin operators $\vec{\tau }$, but they 
now contain off-diagonal 
spin-flip terms.

It is convenient at this stage to examine the off-diagonal structure
of the $\bPhi_{\Gamma \bk }$. These two dimensional matrices are
proportional to unitary matrices, and take the form
\[
\bPhi_{\Gamma\bk}=\phi_{\Gamma\bk}{\cal U}_{\Gamma \bk }
\]
where $\phi_{\Gamma_{\bk }}$ is a scalar and ${\cal U}_{\Gamma\bk }$
is a two-dimensional unitary matrix. This matrix defines
the interconversion between Bloch states and spin-orbit coupled Wannier
states.  When an electron
``enters'' the Kondo singlet, its spin quantization axis is
rotated according to the matrix ${\cal U}_{\Gamma \bk }$.
When it
leaves the ion in the same channel, this rotation process is undone,
and the net hybridization matrix
\[
\bPhi\dg_{\Gamma\bk}
\bPhi_ {\Gamma\bk} = (\phi_{\Gamma\bk })^{2} \ {\bf 1}
\]
is spin-diagonal. However, in the presence of composite pairing 
an  incoming electron
in channel $1$ can Andreev scatter from
the ion as a hole in channel $2$. This leads to a net
rotation of the spin quantization axis 
through an angle $\zeta_\bk$ about an axis ${\bf n}_{\bk }$
that both depend on the
location on the Fermi surface, as follows
\[
\bPhi\dg _{2\bk }\cdot \bPhi_{1\bk} = \phi_{2\bk }\phi_{1\bk } \bigl [
c_{\bk } + i s_{\bk } ( {\bf n}_{\bk }\cdot {\bsig })\bigr ]
\]
where $c_{\bk }= \cos(\zeta_\bk/2)$, $s_{\bk }= \sin(\zeta_{\bk}/2)$. 

Armed with this information, we now continue to examine the resonant
scattering off the composite-paired Kondo singlet. When we expand the
self energy, we obtain a normal and Andreev component, given by 
\bea
\Sigma(\kappa)=\Sigma_N(\kappa)+\Sigma_A(\kappa).
\eea
where 
\begin{eqnarray}\label{l}
\Sigma_N(\kappa)&=& \frac{v_{1 {\bf k}}^2}{ \omega - \lambda \tau_3}
+
\frac{v_{2 {\bf k}}^2}{ \omega + \lambda \tau_3}\cr
&=& \frac{1}{\omega^{2}-\lambda^{2}}\left[\omega (
v_{1\bk}^{2}+v_{2\bk }^{2}
)+ \lambda (v_{1\bk}^{2}-v_{2\bk }^{2}
)\tau_{3} \right]
\end{eqnarray}
and  we denote
$v_{1\bk } = v_{1}\phi_{1\bk}  ,\ 
v_{2\bk }= \Delta_{2}\phi_{2\bk}$.
By contrast, the Andreev terms take the form
\begin{eqnarray}\label{l}
\Sigma_{A} (\kappa)&=& 
\left(\frac{i \omega\tau_{2}-\lambda
\tau_{1}}{(\omega^{2}-\lambda^{2})}\Phi \dg_{2\bk}\Phi_{1\bk }
+ 
{\rm
H.c} \right)
\cr
&=& 
\frac{2v_{1\bk }v_{2\bk }}{(\omega^{2}-\lambda^{2})}\left[-
\lambda c_{\bk }\tau_{1}+ \omega s_{\bk } ({{\bf n}_{\bk }\cdot \bsig })\tau_{2} \right]
\end{eqnarray}

Notice how the Andreev scattering contains two terms:
\begin{itemize}
\item a scalar term $\frac{2v_{1\bk }v_{2\bk }}{(
\lambda^{2}-\omega^{2})}
\lambda c_{\bk }\tau_{1}$  
that is finite at the Fermi energy ($\omega=0$),
with gap symmetry of the form
\[
\Delta_{\bk }\propto {\rm Tr \Phi\dg _{2\bk }\Phi_{1\bk }} \sim \phi_{1\bk }\phi_{2\bk }c_{\bk }
\]

\item a ``triplet'' term $-\omega\frac{2v_{1\bk }v_{2\bk
}}{(\lambda^{2}-\omega^{2}
)}\left[ s_{\bk } ({{\bf n}_{\bk }\cdot \bsig })\tau_{2} \right]
$ which is {\sl odd
} in frequency and vanishes on the Fermi surface. 

\end{itemize}
In practice, the nodes of the pair wavefunction are dominated by the
symmetry of the function  $c_{\bk }$.  When an electron Andreev
reflects through one hybridization channel into the other, it acquires
orbital
angular momentum.  For example, the ``up'' states of the 
$\Gamma_{7}^{+}\sim \vert  -3/2\rangle $ and $\Gamma_{7}^{-}\sim \vert
+ 5/2\rangle $ differ by $l=4 $ units of angular momentum, so the 
resulting gap has the symmetry of an $l=4$ spherical harmonic, or 
$g-$ wave symmetry.  By contrast, the up states of the
$\Gamma_{7}^{+}\sim \vert -3/2\rangle $ and $\Gamma_{6}\sim \vert +
1/2\rangle $ differ by $l=2$ units of angular momentum, and the
resulting gap has the symmetry of an $l=2$ spherical harmonic, or
$d$- wave symmetry, as shown below:

\begin{figure}[tbp]
\includegraphics[width=5in]{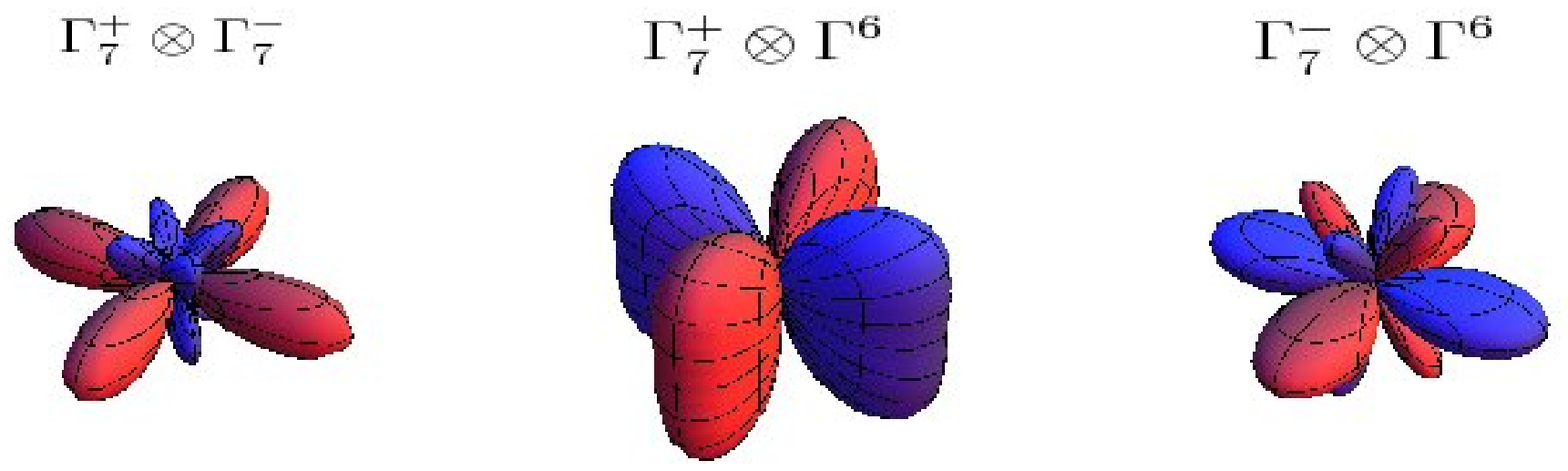}
\caption{Showing the three possible gap functions: $\Gamma^{+}_7 \otimes \Gamma^{-}_7$, $\Gamma^{+}_7 \otimes \Gamma_{6}$, and $\Gamma_{7}^{-} \otimes \Gamma_6$ for mixing angle $\beta  = \pi/10$.  Positive nodes are colored red, while negative nodes are blue. $\Gamma^{+}_7 \otimes \Gamma^{-}_7$ is g-wave(l = 4), while $\Gamma^{\pm}_7 \otimes \Gamma_{6}$ are both d-wave(l=2).}
\label{Fig_Gaps}
\end{figure}

\subsection{Dispersion in the presence of strong spin-orbit coupling }\label{mfso}

To develop a mean-field theory in the presence of spin-orbit
scattering, we need to diagonalize the the conduction electron Green's
function. The eigenvalues are determined by the condition
\[
{\rm det}[\omega \underline{1}- {\cal H} (\bk )]=0
\]
If we integrate out the f-electrons, this becomes
this becomes
\[
{\rm det}[\omega \underline{1}- {\cal H} (\bk )]= 
(\omega^{2}-\lambda^{2})^{2}
{\rm det}[{\cal G} (\kappa)^{-1}]= 
(\omega^{2}-\lambda^{2})^{2}
{\rm det}[\omega - \epsilon_{\bk }\tau_{3}- \Sigma (\kappa)]
\]
Now since $\Sigma (\kappa)\propto \frac{1}{\omega^{2}-\lambda^{2}}$,
it is convenient to factor this term out of the determinant, so that
\[
{\rm det}[\omega \underline{1}- {\cal H} (\bk )]= 
(\omega^{2}-\lambda^{2})^{-2}
{\rm det}[
(\omega^{2}-\lambda^{2}) {\cal G} (\kappa)^{-1}
]
\]
Now 
\begin{equation}\label{}
(\omega^{2}-\lambda^{2}) {\cal G} (\kappa)^{-1} = 
\omega ( A +D (\bsig \cdot {\bf n}_{\bk })\tau_{2})
- B \tau_{3} +
C\tau_{1}
\end{equation}
where
\begin{eqnarray}\label{l}
A&=& \omega^{2}- \lambda^{2}- v_{\bk +}^{2}\cr
B &=& \epsilon_{\bk } (\omega^{2}-\lambda^{2})+ \lambda v_{\bk -}^{2}\cr
C&=& 2 \lambda v_{1\bk }v_{2\bk } c_{\bk }\cr
D&=& 2 v_{1\bk }v_{2\bk }s_{\bk }
\end{eqnarray}
and $v_{\bk \pm }^{2}= v_{1\bk }^{2}\pm v_{2\bk }^{2}$.

Now if we project the Hamiltonian into  states  where
$(\bf n_{\bk}\cdot \bsig)= \pm 1$, we can replace $A+D (\bsig \cdot
{\bf n}_{\bk })\tau_{2}\rightarrow  A \pm D\tau_{2}$, i.e
\begin{eqnarray}\label{l}
{\rm det}[\omega \underline{1}-{\cal H} (\bk)]&=&
\prod_{\pm}
\frac{
{\rm det}[\omega ( A \pm D \tau_{2})
- B \tau_{3} +C\tau_{1}  ]}{(\omega^{2}-\lambda^{2})
}
\end{eqnarray}
The presence of the $\omega^{2}-\lambda^{2}$ terms in the denominator 
results from
integrating out the f-electrons. In actual fact, there are no zeroes
of the determinant at $\omega = \pm \lambda$, and the 
the $\omega^{2}-\lambda^{2}$ denominators in these expressions act to factor
out the false zeros $\omega= \pm \lambda $ in the numerator that have
been introduced by integrating out the f-electrons. If 
now expand the numerator:
\begin{eqnarray}\label{expandit}
{\rm det}[\omega ( A \pm D \tau_{2})
- B \tau_{3} +C\tau_{1}  ]
&=& [\omega^{2 }A^{2}- \omega^{2 }D^{2}
- B^{2} - C^{2}]\cr
= \omega^{2}\left[(\omega^{2}-\lambda^{2}- v_{+}^{2})^{2}- (2 v_{1\bk
}v_{2\bk }s_{\bk })^{2}\right]
&-& \left[\epsilon_{\bk } (\omega^{2}-\lambda^{2})+
\lambda v_{-}^{2}\right]^{2}
- \left[2 \lambda v_{1\bk }v_{2\bk }c_{\bk }\right]^{2}.
\end{eqnarray}
Notice that we get the same result for both $\pm D$. 
Now we know that there is a factor $(\omega^{2}-\lambda^{2})$ in this
expression, so we can write
\begin{eqnarray}\label{l}
{\rm det}[\omega ( A \pm D \tau_{2})
- B \tau_{3} +C\tau_{1}  ]
&=&(\omega^{2}-\lambda^{2})
\left[\omega^{4}- 2 \omega^{2} \alpha_{\bk }+ \gamma^{2}_{\bk } \right]
\end{eqnarray}
By a direct expansion of this expression and a  comparison of terms
with (\ref{expandit}), we are able to confirm that this
factorization works, with 
\begin{eqnarray}\label{betanot}
\alpha_{\bk }&=& v_{\bk +}^{2}+ \frac{1}{2} (\lambda^{2}+\epsilon_{\bk
}^{2})\cr
\gamma^{2}_{\bk }&=& (\epsilon_{\bk }\lambda-v_{\bk -}^{2})^{2}+ (2v_{1\bk }v_{2\bk }c_{\bk} )^{2}
\end{eqnarray}
Thus
\begin{equation}\label{gammanot}
{\rm det}[\omega \underline{1}- {\cal H} (\bk )]= \left[\omega^{4}- 2 \omega^{2} \alpha_{\bk }+ \gamma^{2}_{\bk } \right]^{2}
\end{equation}
The surviving yet crucial effect of the spin-flip scattering is 
entirely contained in the $c_{\bk }$ factor in $\gamma_{\bk }$.
The Boguilubov quasiparticles in the composite paired state preserve their
Kramer's degeneracy, with dispersion given by 
\[
\omega_{\bk\pm}=
\sqrt{\alpha_\bk\pm(\alpha_\bk^2-\gamma_\bk^2)^{1/2}},
\]
as described in section (\ref{mftheorysec}). 

\subsection{Crystal Fields determine the gap symmetry.}\label{}

Here we calculate the form factors for the two-channel
Kondo model in a tetragonal crystal field environment. 
In a tetragonal
crystal field environment, the Kramer's doublets are given by
$\vert \Gamma \pm \rangle = \vert  j m\rangle \langle j m \vert
\Gamma \pm \rangle $, ($j=5/2$), where from (\ref{xtalfields})
\begin{eqnarray}\label{xtalfields2}
\begin{array}{lccl}
\Gamma_{6}:&\qquad f\dg_{\Gamma_{6}\pm} &=& \vert {\pm 1/2}\rangle \cr
\Gamma_{7}^{+}:&\qquad f\dg_{\Gamma_{7}^{+}\pm} &=& \cos \beta  \vert {\mp 3/2}\rangle +
\sin \beta \vert  {\pm 5/2}\rangle \cr
\Gamma_{7}^{-}:&\qquad f\dg_{\Gamma_{7}^{-}\pm} &=& \sin \beta  \vert
{\mp 3/2}\rangle -\cos \beta \vert  {\pm 5/2}\rangle 
\end{array}
\end{eqnarray}
The matrices representing these crystal field states are then
\begin{eqnarray}\label{overlap1}
&& {\textstyle\ m:\  \frac{5}{2} \ 
\qquad \ 
\ \hskip 0.03in -\frac{5}{2}}  \ \   \hskip 0.15in \alpha :
\cr
\langle \Gamma_{6}\alpha \vert j m \rangle  &=& \ \ \ \ \hskip 0.03in \left(\begin{matrix}
0 & 0 & 1 & 0 & 0 & 0\cr
0 & 0 & 0 & 1 & 0 & 0 
\end{matrix}
 \right),  \begin{matrix} + \cr - \end{matrix} 
\cr
\langle \Gamma_{7^{+}}\alpha \vert j m \rangle  &=&
 \left(\begin{matrix}
\sin(\beta) & 0 & \ \ 0 &\ \  0 & \cos(\beta)& 0  \cr
0 & \cos(\beta)  &\  \ 0 &\  \ 0 & 0 &
\sin(\beta) 
\end{matrix}
 \right), 
\cr 
\langle \Gamma_{7^{-}}\alpha \vert j m \rangle  &=&  \left(\begin{matrix}
- \cos(\beta) 
& 0 & \ \ 
0 & \ \  0 & \sin(\beta) & 0  \cr
0 & \sin(\beta) 
 & \ \  0 &  \ \ 0 &  0 & \cos(\beta) 
\end{matrix}
 \right). 
\end{eqnarray}
while the overlap of the Bloch states with the spin-orbit coupled
Wannier states $\vert j m\rangle $ ($j=5/2$, $m\in [-j,j]$) is given
by  
\begin{equation}\label{overlap2}
\langle j m \vert \bk  \sigma  \rangle = 
\sigma
\sqrt{\frac{\frac{7}{2}-m \sigma }{7}}
Y^{3}_{m-\frac{1}{2}\sigma } (\hat  \bk )  = 
\left(\begin{matrix} 
\sqrt{\frac{1}{7}}Y^{3}_{2} (\bk )& 
-\sqrt{\frac{6}{7}}Y^{3}_{3} (\bk )\cr
\vdots & \vdots \cr
\sqrt{\frac{6}{7}}Y^{3}_{-3} (\bk )& 
-\sqrt{\frac{1}{7}}Y^{3}_{-2} (\bk )
\end{matrix}
 \right)
.
\end{equation}
where we have used the Clebsch Gordon coefficient $\langle jm \vert l m-\frac{1}{2}\sigma,
\frac{1}{2}\ \frac{1}{2}\sigma \rangle =  \sigma
\sqrt{\frac{\frac{7}{2}-m \sigma }{7}}
$,  ($\sigma = \pm 1$). 
We obtain the form factors of the Wannier states by multiplying
matrices from (\ref{overlap1}) with the matrix elements (\ref{overlap1}):
\begin{equation}\label{}
[\Phi_{\Gamma \bk }]_{\alpha \sigma }= \sum_m \langle \Gamma \alpha
\vert j m \rangle \langle j m \vert \bk \sigma \rangle 
\end{equation}
The form factors are then 
given by
\begin{equation}\label{nn}
\Phi_{\Gamma_{7}^{+}\bk }= 
\frac{1}{\sqrt{7}}
\left[
\begin{matrix}
{\sqrt{5} c Y^{3}_{-2}( \hat \bk )+s
Y^{3}_{2}( \hat \bk )}
 & 
-\sqrt{6} s Y^{3}_{3}( \hat \bk )-\sqrt{2}c
Y^{3}_{-1}( \hat \bk )
\cr
\sqrt{6} s Y^{3}_{-3}( \hat \bk )
+\sqrt{2}c Y^{3}_{1}( \hat \bk )
 & \
-\sqrt{5} c  Y^{3}_{2}( \hat \bk )-s Y^{3}_{-2}( \hat \bk )
\end{matrix}
\right],
\end{equation}
\begin{equation}\label{nn}
\Phi_{\Gamma_{7}^{-}\bk }= 
\frac{1}{\sqrt{7}}
\left[
\begin{matrix}
{\sqrt{5} s Y^{3}_{-2}( \hat \bk )-c
Y^{3}_{2}( \hat \bk )}
 & 
\sqrt{6} c Y^{3}_{3}( \hat \bk )-\sqrt{2}s
Y^{3}_{-1}( \hat \bk )
\cr
- \sqrt{6} c Y^{3}_{-3}( \hat \bk )
+
\sqrt{2}s Y^{3}_{1}( \hat \bk \
)
 & \
-\sqrt{5} s  Y^{3}_{2}( \hat \bk )
+c Y^{3}_{-2}( \hat \bk )
\end{matrix}
\right]
\end{equation}
\[
\Phi_{\Gamma_{6}\bk } = \frac{1}{\sqrt{7}}\left[
\begin{matrix}

\sqrt{3}Y^{3}_{0} (\hat  \bk ) 
& 
-\sqrt{4}Y^{3}_{1} (\hat  \bk ) \cr
\sqrt{4}Y^{3}_{-1} (\hat  \bk ) &
- \sqrt{3}Y^{3}_{0} (\hat  \bk )
\end{matrix}
 \right]
\]
where we use the shorthand $c\equiv \cos(\beta)$, $s \equiv
\sin(\beta)$ to denote the cosine and sine of the mixing angle. 
Each of these functions is time-reversal invariant, namely they satisfy
\[
\epsilon (\Phi\dg _{\Gamma - \bk })^{T}\epsilon^{T} = \Phi_{\Gamma \bk }
\]
There are basically two symmetry classes of superconductor that are possible in
our model, one formed from $\Gamma_{7}\otimes \Gamma_{6}$, the other
formed from $\Gamma_{7}^{+}\otimes \Gamma_{7}^{-}$ (Fig. \ref{Fig_Gaps}). The latter is
argued to be preferably, because it is these two states that have the
maximum overlap with nearby ligand atoms. The form factors
$\phi_{\Gamma \bk }$ are determined from
\begin{eqnarray}\label{l}
(\phi_{1\bk })^{2}&=&\frac{1}{2}{\rm Tr}[\Phi\dg_{1\bk }\Phi_{1\bk }]\cr
(\phi_{2\bk })^{2}&=&\frac{1}{2}{\rm Tr}[\Phi\dg_{2\bk }\Phi_{2\bk }]
\end{eqnarray}
For general $\beta $ , these are form-factors with point nodes along
the c-axis, but no lines of nodes.
The gap function is however determined from the symmetry of
\[
\phi_{1\bk }\phi_{2\bk}c_{\bk } = 
\frac{v_{1}\Delta_{2}}{4}{\rm Tr}[\Phi\dg _{2\bk }\Phi _{1\bk }+ {\rm H. c.}]
\]
In our model, we have chosen $\Gamma_{1}\equiv \Gamma_{7}^{+}$
corresponding to the out-of-plane ligand atoms and $\Gamma_{2}\equiv
\Gamma_{7}^{-}$ corresponding to the in-plane ligand atoms.
In this case, when we expand this function in detail, we find it has the form
\[
\phi_{1\bk }\phi_{2\bk}c_{\bk } = \cos(2\beta)\Delta_{g1}(\bk)  - \sin
(2 \beta )\Delta_{g2} (\bk )
\]
where $\Delta_{g1}$ and $\Delta_{g2}$ are g-wave gap functions of the form
\[
\Delta_{g1} (\bk  ) = \frac{\sqrt{5}}{16 \pi}\cos(4\phi)\sin^2[\theta]
\]
and
\[
\Delta_{g2} (\bk )= \frac{3}{16\pi}\sin^2(\theta) (3 + \cos (2 \theta )).
\]
For small $\beta $, the gap is dominated by $\Delta_{g1} (\bk )$

\section{Frustrated Magnetism}

As a test of symplectic-$N$, we would like to return to the origins of $SP(N)$, initially developed to treat antiferromagnetism on frustrated lattices\cite{readsachdev91}.   $SP(N)$ describes antiferromagnetism, or the formation of valence bonds well, but cannot handle ferromagnetism, the fluctuations of those bonds.  This weakness can already be seen in the simplest model in frustrated magnetism, the $J_{1}-J_{2}$ Heisenberg model on a square lattice
\cite{doucot}
\begin{equation}
H = J_{1}\sum_{\bx , \mu}  \vec{S}_\bx  \cdot \vec{S}_{\bx +\mu}+ 
J_{2}\sum_{\bx ,\mu'}\quad \vec{S}_{\bx }\cdot \vec{S}_{\bx +\mu'},
\end{equation}
where $J_{1}$ and $J_{2}$ are the first and next nearest neigbor
couplings.  
\fg{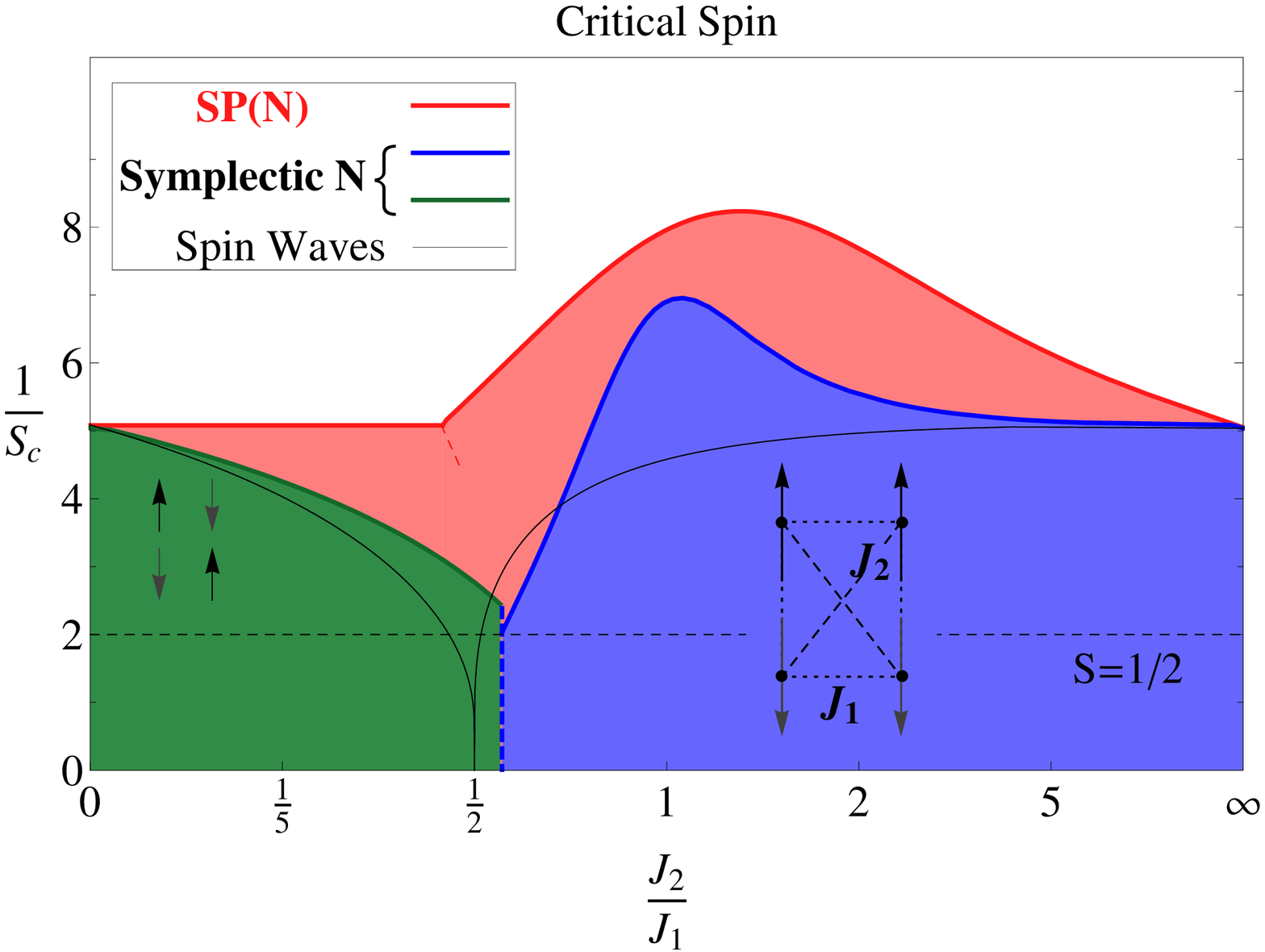}{Sc}{The $J_1-J_2$ Heisenberg model: We
compare the critical spin  $S_c = \left(\frac{n_{b}}{N}\right)_c$,
below which there is no long range order in the ground state, calculated within
$SP(N)$ (bold red line), symplectic-$N$ (blue and green lines), and spin wave
theory\cite{chandradoucot} (thin black line).  $J_1$ and $J_2$ are nearest and next nearest
neighbor antiferromagnetic bonds, as shown in the figure.  For small
$J_2/J_1$, the spins configurations are staggered, 
while for large $J_2/J_1$, the
ground state breaks lattice symmetry to develop 
collinear order as shown in the figure.  $SP(N)$ (bold red line)
tends to overstabilize the long range ordered phases, most
dramatically on the one sublattice side, where the critical spin is
independent of the strength $J_{2}$ of the 
frustrating diagonal bonds\cite{oleg}.  Symplectic-$N$
restores the frustration-induced fluctuations by treating both
ferromagnetic and antiferromagnetic bonds, on equal footing, which
corrects this  overstabilization.}

For large $J_{1}/J_2$, the ground state is the typical N\'eel state,
where diagonal spins are ferromagnetically aligned(see Fig \ref{Sc} inset).
As $J_2$ increases, these diagonal bonds become increasingly
frustrated, destabilizing the long range order, which requires higher
and higher spin, $S_c$ for magnetic order. 

At large $J_{2}/J_1$,
this model describes two interpenetrating N\' eel lattices that are
classically decoupled. When fluctuations are included, a biquadratic
interaction locks the two sublattices together in a collinear
configuration\cite{ccl}(see Fig \ref{Sc}).  The nearest neighbor bond, $J_1$,
initially stabilizes the collinear state, but, as the frustration
increases, the long range order is eventually destabilized.  

In quantum magnetism, one generally uses a Schwinger boson spin
representation. The $SP (N)$ approach introduced by Read and Sachdev
\cite{readsachdev91long} decouples the antiferromagnetic interaction
in the following manner
\[
H_{I}= -\frac{J}{N} B\dg_{12}B_{12}
\quad \qquad \qquad \qquad \qquad \qquad \qquad SP (N).
\]
where  $B\dg _{21} 
=\sum_{\sigma }\tilde{\sigma } b\dg_{2\sigma }b\dg _{1-\sigma }
$ creates a valence bond $(2,1)$ between sites one and two. 
The $SP(N)$ approach
captures the competition between first and next nearest neighbor links
for valence bonds\cite{oleg}, but misses the frustrating effect of the
ferromagnetic bonds, resulting in overstabilization of the collinear
state\cite{readsachdev91long}. If we decompose $H_{I}$ in terms of
spin generators, we find it contains a mix of ``spins'' and ``dipoles''
\begin{equation}\label{}
H_{I}= \frac{J}{2N} ( \vec{ S}_{1}\cdot \vec{S}_{2}- {\vec{\cal  P}}_{1}
\cdot {\vec{\cal  P}}_{2}),\quad \qquad \qquad \qquad  \underline{SP (N)}.
\end{equation}
This  inadvertent inclusion of 
dipoles with a negative, i.e ferromagnetic sign, tends to 
cancelling out the frustrating effect of ferromagnetic bonds.  For
instance, in 
the $J_{1}-J_{2}$ model, the critical spin for developing
long-range  antiferromagnetism is artificially 
independent of the frustration in the 
the $SP (N)$ mean-field theory\cite{oleg} (Fig. \ref{Sc}.). 

In the symplectic- $N$ approach, we decouple the interaction
exclusively in terms of the generators of $SP (N)$ 
$S_{\alpha \beta } = b\dg_{\alpha }b_{\beta }- \tilde{ \alpha }\tilde{\beta }
b\dg_{- \beta }b_{\alpha}$.  Using the explicit form
of the symplectic spins, we find the Heisenberg interaction 
decouples into two terms
\begin{equation}\label{spinham}
H_{I}= \frac{J}{N}\vec{S}_{1}\cdot\vec{S}_{2} =
-\frac{J}{N}\left[ B\dg_{21}B_{21}- A\dg_{21}A_{21}\right],\qquad
\qquad  {\hbox{\underline{Symplectic $N$.}}}
\end{equation}
where $A_{21}=
\sum_{\sigma }b\dg_{2\sigma }b_{1\sigma }$. 
The second term in this interaction 
describes \emph{ferromagnetic}
correlations, which were absent in the original applications.  
The operators $A_{21}$ and $A\dg _{21}$
``resonate''  a valence bond linked to a third
site, between sites one and two:  
$(1,3)\rightleftharpoons (2,3)$.
When the bond
resonates between sites 1 and 2, the amplitude for singlets to
form between the two sites is reduced, giving rise to a 
ferromagnetic correlation between sites 1 and 2. 
The exclusion of dipole spins requires that we treat these two terms
in  equal measure.  

When we carry out 
a Hubbard Stratonovich factorization of the Heisenberg interaction (\ref{spinham}), it
separates into two amplitudes $h$ and 
 $\Delta $ describing bond resonance and condensation respectively
\begin{equation}\label{HStransform}
J\vec S_{1} \cdot \vec{S}_{2}  =
\left(b\dg_{2\sigma}, \tilde{\sigma} b_{2\ - \sigma }\right)
\left[\begin{matrix}
h &\Delta \cr
\bar \Delta & \bar h
\end{matrix} \right]
\left(\begin{matrix}
b_{1\sigma}\cr
\tilde{\sigma }b\dg_{1\ -\sigma }
\end{matrix} \right)
+
\frac{N}{J}(|\Delta|^2 - |h|^2).
\end{equation}
This kind of decoupling scheme was first proposed by Ceccatto, Gazza and Trumper\cite{ceccatto} for $SU(2)$ spins, where it is one of many alternative decoupling procedures.    In  symplectic-$N$, it is the unique form preserving the time reversal parity of the spins.
Now we would like to see if and when the ferromagnetic $h$ bonds develop and what effect they have on the physics.  To do this we examine the action,
\begin{equation}
N {\cal S}[b, \Delta,h,\lambda] = \int_0^{\beta} d\tau \sum_i [\sum_\sigma b^\dagger_{i\sigma}(\partial_\tau - \lambda_i) b_{i\sigma} + \lambda_i NS] + \sum_{(ij)} J_{ij}\vec{S}_i \cdot \vec{S}_j
\end{equation}
with $J_{ij}\vec{S}_i \cdot \vec{S}_j$ given above.  We assume all bond fields are uniform and static, depending only on $i-j$.  The constraint, $n_b = NS$ is enforced by the Lagrange multiplier $\lambda_i$.  As the action is quadratic in the Schwinger bosons, they can easily be integrated out to find the mean field Free energy:
\begin{equation}
\label{free}
\frac{F_{MF}}{N \mathcal{N}} = \frac{1}{\mathcal{N}}\sum_k \log[ 2 \sinh \frac{\beta \omega_k}{2}] + \frac{1}{\mathcal{N}}\sum_{(i,j)} \frac{ \bar{\Delta}_{ij}\Delta_{ij}-\bar{h}_{ij} 
h_{ij}}{J_{ij}} -\lambda(2S+1)
\end{equation}
where $(i,j)$ is a pair of sites with nonzero $J_{ij}$, $\mathcal{N}$ is the number of sites,  $ \omega_k = \sqrt{|\lambda-h_k|^2-|\Delta_k|^2}$, and $h_k$ and $\Delta_k$ are the Fourier transforms of $h_{ij}$ and $\Delta_{ij}$.  
The N\'eel state is described by $\Delta$'s along the nearest neighbor bonds, and induced $h_d$ along the diagonal bonds for finite $J_2$. 
\begin{equation}
\omega_k^{Neel} = \sqrt{(\lambda + 4 h_d c_x c_y)^2 - 4 \Delta^2(s_x+s_y)^2)},
\end{equation}
where $c_l = \cos k_l a$ and $s_l = \sin k_l a$.
For large $J_2$, the classical decoupled state consists of $\Delta_d$ along the diagonal bonds, where the magnitude of $\Delta_d$ is the same on both sublattices, but the phase between the two is free.  When fluctuations are introduced, the lattice symmetry is broken by the collinear order; we choose the phase so the collinear state is antiferromagnetic along $\hat y$($\Delta_y$), which induces ferromagnetism along $\hat x$($h_x$), giving the dispersion,
\begin{equation}
\omega_k^{Ising} = \sqrt{(\lambda + 2 h_x c_x)^2 -  (4 \Delta_dc_x s_y+2 \Delta_y s_y)^2)}
\end{equation}

The values of $h$, $\Delta$, and $\lambda$ are chosen to minimize the Free energy using the mean field equations: $\partial F/\partial \lambda = 0$, $\partial F/\partial h = 0$ and $\partial F/\partial \Delta = 0$, for all $h$'s and $\Delta$'s in the particular problem.  $SP(N)$ proceeds similarly except that the $h$'s are fixed to be zero.

We are interested in the zero temperature case, where in two dimensions, the Schwinger bosons can condense when the gap in the spectrum vanishes, signaling the onset of long range magnetic order.  As we are primarily interested in the spin above which the system orders magnetically, $S_c$, we only consider the unpopulated condensate.

The minimum gap in the spectrum is fixed at $(\pm \pi/2,\pm \pi/2)$ and $(0, \pm \pi/2)$ for the N\'eel and Ising states, respectively.  Setting that gap to zero gives us an algebraic relation between the parameters supplementing the mean field equations.  After the parameters are determined from $\partial F/\partial h$ and $\partial F/\partial \Delta$, $\partial F/\partial \lambda = 0$ can be used to find the critical spin,
\begin{equation}
S_c + \frac{1}{2} = \frac{1}{2}\int_k \frac{\lambda + 2 h_k}{\omega_k}.
\end{equation}

Results from these calculations for both symplectic-$N$ and $SP(N)$ are shown in Fig {\ref Sc}, along with a comparision to spin wave theory.  $SP(N)$ drastically overestimates the stability of the ordered states, which is corrected by the ferromagnetic bonds included in symplectic-$N$.  For both large $N$ theories, the regions of N\'eel and Ising order overlap, indicating a first order transition, while spin wave theory predicts a second order transition for all $S$, with an intervening region of spin liquid.  However, when higher order corrections are taken into account, modified spin wave theory gives exactly the results of symplectic-$N$\cite{wu}.

We can draw an analogy between frustrated magnetism and heavy fermion superconductivity, in which previous large $N$ techniques were able to treat only one of two possible phenomena, ferromagnetism and antiferromagnetism in the case of frustrated magnetism, Fermi liquid physics and superconductivity in heavy fermion superconductors.  In both situations, symplectic-$N$ enables simultaneous and equivalent treatment of both phenomena and significantly improves upon the previous results.

\end{document}